\documentclass[11pt,letter]{article}
\usepackage{jcappub}
\usepackage{aas_macros}

\usepackage{graphicx}
\usepackage{bm}
\usepackage{url}

\newcommand\be{\begin{equation}}
\newcommand\ee{\end{equation}}

\newcommand\bea{\begin{eqnarray}}
\newcommand\eea{\end{eqnarray}}

\newcommand\eq[1]{eq.~(\ref{eq:#1})} 
\newcommand\fig[1]{fig.~(\ref{fig:#1})} 

\newcommand\nn{ \notag \\}

\newcommand{\diff}{\mathrm{d}}
\newcommand{\mchi}{m_\chi}
\newcommand{\vrel}{v_\mathrm{rel}}
\newcommand{\vrelt}{\bm{v}_\mathrm{rel}}

\newcommand\pvx[2][{}]{P_{\bm{x}} \left(\bm{#2}_{#1} \right)}

\newcommand{\nchi}{n_{\chi}}

\newcommand{\rvir}{r_\mathrm{vir}}
\newcommand{\mvir}{M_\mathrm{vir}}
\newcommand{\rhot}{\tilde{\rho}}
\newcommand{\epst}{\tilde{\epsilon}}
\newcommand{\psit}{\tilde{\psi}}

\title{The impact of the phase-space density on the indirect detection
of dark matter}
\author{Francesc Ferrer,}
\author{Daniel R. Hunter}
\affiliation{Physics Department and McDonnell Center for the Space Sciences\\ 
Washington University, St Louis, MO 63130, USA}
\abstract{
We study the indirect detection of dark matter
when the local dark matter velocity distribution 
depends upon position, as expected for the Milky Way and 
its dwarf spheroidal satellites, and the annihilation cross-section
is not purely s-wave.
Using a phase-space distribution consistent with the dark matter
density profile, we present estimates of cosmic and gamma-ray fluxes from 
dark matter annihilations. The expectations for the indirect detection
of dark matter can differ significantly from the usual
calculation that assumes that the velocity of the dark matter particles
follows a Maxwell-Boltzmann distribution. 
}

\begin{document}
\maketitle

\section{Introduction}

Besides dark matter's gravitational effects, which establishes it as the second
largest contributor to the energy budget of the Universe, very little is
known about its fundamental nature. A popular, and well-motivated, candidate 
for composing the dark matter is a weakly interacting massive particle (WIMP).
With mass and interactions at the electro-weak energy scale, 
WIMPs naturally obtain a thermal relic density in the range 
required by cosmological observations. The foremost example of WIMP is
the lightest neutral supersymmetric particle, the 
neutralino~\cite{Jungman:1995df}.

Importantly, weak-scale interactions with Standard Model particles make
it possible, in principle, to detect the presence of WIMPs. Direct detection
experiments monitor the recoil of nuclei that might be elastically 
scattered by dark
matter particles. On the other hand, WIMPs can self-annihilate and 
produce gamma-rays, neutrinos, or other Standard Model particles.
This is the basis for the indirect
detection of dark matter~\cite{Jungman:1995df,Bertone:2004pz}.

Over the past few years, direct detection experiments have attained the 
sensitivity to probe the parameter space of the simplest
supersymmetric extensions of the Standard Model (SM), and
next-generation multi-ton experiments~\cite{Newstead:2013pea} are being 
planned to test less constrained extensions~\cite{Ellis:2009ai,Freese:2012xd}. 
Even though, motivated by theoretical expectations on the neutralino, 
experimental efforts have focused on the DM mass range 
$100\;{\rm GeV} \lesssim m_\chi \lesssim 1 \;{\rm TeV}$,
several experiments have reported signals that can be 
attributed to low-mass dark matter particles, 
$m_\chi 
\lesssim 10\;{\rm GeV}$~\cite{Bernabei:2010mq,Aalseth:2012if,Agnese:2013rvf}.
The strong tension of this interpretation with the constraints
set by the null results of other low-background experiments, such as
XENON100 and CDMS-Ge~\cite{Aprile:2012nq,Ahmed:2010wy}, 
can be eased, although not completely 
eliminated~\cite{Frandsen:2013cna}, by considering
dark matter particles with non-standard 
interactions~\cite{TuckerSmith:2001hy,Kurylov:2003ra,Feng:2011vu} and
systematic errors in the determination of the detector response at low
recoil energies~\cite{Savage:2010tg,Kelso:2011gd}.
The interpretation of the data depends also on the assumed
velocity distribution of the DM 
halo~\cite{McCabe:2010zh,Frandsen:2011gi,Green:2011bv,Fairbairn:2012zs},
and this has prompted investigators to consider more realistic
scenarios than the Standard Halo with a
Maxwell-Boltzmann 
distribution~\cite{Ullio:2000bf,Lisanti:2010qx,Vogelsberger:2008qb,Belli:2002yt,Catena:2011kv,Fox:2010bu}.

Similarly, several tantalizing observations of cosmic and gamma-ray
fluxes have been linked to the annihilation or decay of DM particles. 
The bright $511$ keV line emission from the bulge of the Galaxy detected
by the SPI spectrometer on the INTEGRAL satellite~\cite{Knodlseder:2003sv},
the excesses of microwaves and gamma-rays in the inner Galaxy revealed
by the WMAP and Fermi satellites~\cite{Su:2010qj}, the evidence for
a $130$ GeV spectral line in the Fermi 
data~\cite{Bringmann:2012vr,Weniger:2012tx}, and the rise in the positron
fraction above $10$ GeV observed by PAMELA and 
AMS-02~\cite{Adriani:2008zr,Aguilar:2013qda}, have been attributed to
the effects of
DM~\cite{Boehm:2003bt,Hooper:2003sh,Hooper:2013rwa,Hooper:2011ti,Abazajian:2012pn}.

In addition to spectral features, the fact that the signals in the photon
sector originate in the Galactic center bodes well for the DM interpretation. 
Indeed, the fluxes from DM annihilations depend on the square of the
density, and N-body simulations suggest that DM halos are {\em cusped}
at the center~\cite{Kuhlen:2012ft}. However, DM particles must have 
non-standard interactions to account for these observations. An appealing 
feature of the WIMP miracle is that the cross-section that enters
the predictions for the indirect detection is directly related to the
thermal cross-section, $\sigma v \sim 3 \times 10^{-26} \; {\rm cm^3/s}$,
that explains the observed cosmological abundance of DM. Since the DM 
particles were relativistic at decoupling but are moving at 
non-relativistic speeds in the halo of the Galaxy, a {\em p-wave} 
velocity dependent flux, $\sigma v \sim b v^2$, can result in the
right relic density for a light MeV dark matter particle that annihilates
into non-relativistic $e^\pm$ pairs with $\sigma v \sim 10^{-5}$ pb, giving 
rise to the
observed 511 keV emission from the bulge~\cite{Boehm:2002yz,Boehm:2003hm}.

Similarly, barring the presence of a nearby DM 
clump~\cite{Hooper:2008kv,Pieri:2009je},
the thermal cross-section falls short by about two orders of magnitude
to explain the AMS-02 data. Here, the correct relic abundance and a larger
annihilation cross-section can be reconciled if the annihilation follows
$\sigma v \propto 1/v$. This so-called {\em Sommerfeld enhancement} results
from the exchange of light particles, and is
at the basis for the DM explanation of the rising positron fraction at GeV
energies~\cite{Hisano:2004ds,ArkaniHamed:2008qn,Pospelov:2008jd,Cirelli:2007xd}.
These scenarios face stringent constraints from anti-proton, gamma-ray
and synchrotron 
data, and there is some tension with the higher DM density in the 
galactic center predicted by the steep NFW and Einasto profiles favored by 
N-body simulations~\cite{Donato:2008jk,Bertone:2008xr,Bergstrom:2008ag,Cirelli:2009dv,DeSimone:2013fia}.

In the simplest WIMP models the annihilation flux is mostly {\em s-wave},
and independent of the velocity, $\sigma v \sim a$. As mentioned above,
then the flux from DM annihilations varies through the halo tracking the square
of the density, which is expected to be larger at the center of the halo
or in the dwarf Spheroidal satellites of the Milky Way~\cite{Evans:2003sc}.
When the annihilation is velocity dependent, however, the flux is 
also affected by the distribution of DM particle velocities, which depends
on the location in the halo. For instance, 
for the same density profile, a 
p-wave annihilating DM gives a shallower distribution
of the halo flux~\cite{Ascasibar:2005rw} and results in a different power
spectrum for the diffuse cosmological signal~\cite{Campbell:2011kf}.
In the case of Sommerfeld enhancement, the flux is greatly increased
towards the center of the halo due to smaller DM 
velocities~\cite{Robertson:2009bh}.

In this paper we take a closer look at the distribution of the annihilation
rate of DM particles in the halo for a general velocity dependent
cross-section. Previous works applied the Jeans equation to estimate
the variance in the velocity within the halo, which was then used as a proxy
for the relative velocity of the annihilating DM particles. Although this
is enough for p-wave DM, it does not take into account kurtosis and
other deviations from Gaussianity in the velocity distribution that
might be important for Sommerfeld enhanced and more general models. 
After reviewing the calculation of the DM annihilation rate in 
Sec.~\ref{sec:dmflux}, we list the different DM profiles that we use to
describe galactic and dSph sized haloes in Sec.~\ref{sec:dmprofile}. An
important aspect of our treatment is that we include the effect of baryons
that dominate the gravitational potential in the central regions of the 
Galaxy. For each halo we find a corresponding single particle
velocity distribution using Eddington's formula in 
Sec.~\ref{sec:vdistribution}, which we then use to derive the {\em relative}
velocity distribution. Sec.~\ref{sec:flux} contains our calculation
of the DM annihilation rates, which 
turn out to be larger than previous estimates based on the Jeans equation.
As discussed in Sec.~\ref{sec:discussion}, this aggravates the tension between
an interpretation of the rising positron fraction based on DM annihilations
and the constraints from gamma-ray and synchrotron data.

\section{The dark matter annihilation rate}
\label{sec:dmflux}
%The annihilation of pairs of slowly moving dark matter particles 
%in the Galactic halo, the center of the Sun, or other regions, is the 
%process that governs the yield of energetic neutrinos, anomalous fluxes of 
%cosmic rays, and the emission of high energy $\gamma$-radiation that can 
%be used to constrain, or potentially detect, the presence of particle dark 
%matter.

A pair of dark matter particles, $\chi$, may annihilate into a final
state consisting of Standard Model particles
%,
%\be
%	\chi \chi \rightarrow a + b + \ldots,
%	\label{eq:annreac}
%\ee
%where $a, b = \gamma, \nu, e^\pm, p, \bar{p}, \ldots$,
with a probability per unit time
\be
\diff \Gamma_{2\chi} = \diff \sigma \times \Phi_{2\chi}.
\label{eq:rate}
\ee
Here, $\diff \sigma$ is the differential cross-section for
the annihilation process, and $\Phi_{2\chi}$ 
is the flux of either initial particle at the position of the other one 
defined as the product of the number density $n_\chi$ and the relative velocity
$u_\chi$:
\be
\Phi_{2\chi} = u_\chi \: n_\chi= u_\chi \frac{ \rho_\chi}{m_\chi},
\label{eq:pflux}
\ee
where $\rho_\chi$ is the DM density.
In a frame where the annihilating dark matter particles have four-momenta 
$p_i = \left(E_i, \bm{p}_i\right)$, $i=1,2$, the relative velocity
takes the value
\be
u_\chi = \frac{\sqrt{\left(p_1 \cdot p_2 \right)^2 - \mchi^4}}{E_1 E_2}.
\label{eq:uchi}
\ee
In the center of mass frame, $\bm{p}_1 = -\bm{p}_2 = \bm{p}$, and 
\eq{uchi} reduces to $u_\chi=\left|\frac{\bm{p}}{E_1}-
\frac{-\bm{p}}{E_2}\right| = \left| \bm{v}_1 - \bm{v}_2 \right|$. When
the two annihilating particles are identical, $E_1=E_2$
in the center of mass frame, and $u_\chi=2 \left| \bm{v}\right|$, which
is the magnitude of the relative velocity, $\vrel$.\footnote{However, in
this frame $u_\chi$ is not really a physical velocity, since it can
take values as large as 2 for extremely relativistic particles.}

The density $\nchi$ will depend
on the position in the halo as described below.
On the other hand, the differential cross-section depends 
on the momenta of the initial
and final particles, and on the particle physics model, but not on the
spatial coordinates.\footnote{We do not consider environmental effects, such as
screening mechanisms, that could play a role in the dark energy sector
(see e.g.~\cite{Khoury:2010xi} for a review).}
%In the astrophysical
%setting that we are considering, a fraction of the products of a large
%number of annihilation reactions, \eq{annreac}, in a given volume 
%element, $\diff V$, will arrive at a detector, which will typically 
%count over time, and record the energy, of particles of a given type, 
%$j = \gamma, \nu, e^\pm, p, \bar{p}, \ldots$. These are related to 
%to the weighted sum of \eq{rate} over the possible reactions, $i$,
%with type $j$ particles
%in the final state,
%and averaged over the possible initial configurations of the annihilating
%dark matter particles. 
%More concretely, 
The rate of particles of type $j$ 
that are generated in a volume
element $\diff V$ at the position $\bm{x}$ in the halo, 
containing $n_\chi (\bm{x})
\diff V$ dark matter particles, is:
\be
\frac{\diff ^2 \Gamma}{\diff E\diff V} = 
n_\chi(\bm{x}) \sum_i { {BR_i} \frac{\diff N_{j,i}}{\diff E} 
\langle \sigma_i \times \Phi_{2 \chi} \rangle_{\bm{x}}},
\label{eq:ratev}
\ee
where $BR_i$ is the branching ratio for the reaction $i$, that produces
an average of
$\diff N_{j,i}$ particles of type $j$ with energies between $E$ and $E +
\diff E$, and $\langle . \rangle$ is the average over all possible 
initial kinematic configurations of the DM particles.

Though dark matter particles had relativistic energies in the
early universe~\cite{bernstein}, 
e.g. around the time of decoupling and freeze-out,
the annihilating particles in the weak gravitational
field of the galactic halo are moving
at non-relativistic speeds, $v \sim 10^{-3} c$, and can be
fully described by a Newtonian\footnote{Except in the region
close to the central black hole~\cite{Banados:2009pr,Sadeghian:2013laa}.} 
distribution function (DF) $f$ 
such that $f(\bm{x},\bm{v},t) \diff^3 \bm{x} \diff^3 \bm{v}$ is the 
probability that a particle $\chi$ has phase-space coordinates in the 
given range at time $t$~\cite{binney}.
We will be only concerned with steady-state
systems, and we drop the explicit time dependence in the DF. 
Also, since some popular dark matter density profiles do not have a 
well-defined finite total mass, unless they are truncated, it
is convenient to redefine the DF so that it corresponds to the
mass density in phase-space. 

The dark matter
density in \eq{pflux} can be recovered from the DF by marginalizing
over velocities
\be
\rho_\chi (\bm{x}) \equiv \int{\diff^3 \bm{v} f(\bm{x},\bm{v})}. 
\label{eq:nchi}	
\ee
%where $N = M/m_\chi$ is the total number of dark matter particles
%making up a halo of mass $M$.
The dark matter particles at the point $\bm{x}$ in the halo have
velocities following the distribution
\be
P_{\bm{x}} (\bm{v}) = \frac{f(\bm{x},\bm{v})}{\rho (\bm{x})},
	\label{eq:vchi}
\ee
which is properly normalized to one.
We can then explicitly write the average over initial
configurations in \eq{ratev} as:
\be
\frac{\diff ^2 \Gamma}{\diff E\diff V} = n_\chi \left(\bm{x}\right)
\sum_i{ {BR_i} \frac{\diff N_{j,i}}{\diff E} \int{
		\diff^3 \bm{v}_1 \diff^3 \bm{v}_2 \: \pvx[1]{v} \pvx[2]{v}
		%f(\bm{x},\bm{v}_1) 	f(\bm{x},\bm{v}_2) 
		\: \sigma_i \times 
\Phi_{2 \chi}}}.
\label{eq:avratev}
\ee

We can further simplify this expression in the center of mass frame of
the annihilating particles by introducing the \emph{relative} velocity 
distribution. 
Given the velocity distribution for a single particle as found above,
we can find the relative velocity distribution
in the center of mass frame of the annihilating particles by noting that 
the probability of two particles having velocities 
$\bm{v}_1$ and $\bm{v}_2$ must be equal to the probability of a 
pair of particles having center of mass velocity 
$\bm{v}_\mathrm{cm} = \left(\bm{v}_1 + \bm{v}_2\right)/2$ and 
relative velocity $\vrelt = \bm{v}_1 - \bm{v}_2$, or, in terms of the 
individual velocity distribution, \eq{vchi},
\begin{align}
	\pvx[1]{v} \pvx[2]{v}\diff^3 \bm{v}_1  \diff^3 \bm{v}_2 &=
P_{\bm{x}} \left(\bm{v}_\mathrm{cm} +\vrelt/2 \right) 
P_{\bm{x}} \left(\bm{v}_\mathrm{cm} - \vrelt/2 \right)
\diff^3 \bm{v}_\mathrm{cm} \diff^3 \vrelt \notag \\ &\equiv
P_{\bm{x}, pair}\left(\bm{v}_\mathrm{cm},\bm{v}_\mathrm{rel}\right) 
\diff^3 \bm{v}_\mathrm{cm} \diff^3 \vrelt.
\label{eq:vpair}
\end{align}
Integrating over the center of 
mass velocity, since the annihilation process is invariant under translations,
and writing \eq{rate} in the center of mass frame, we obtain a general expression 
for the production rate of particles of
type $j$:
\be
\frac{\diff ^2 \Gamma}{\diff E\diff V}= n_\chi^2 \left(\bm{x}\right)
\sum_i{ {BR_i} \frac{\diff N_{j,i}}{\diff E} \int{
		\diff^3 \vrelt \: 
		P_{\bm{x}, \mathrm{rel}}\left(\bm{v}_\mathrm{rel}\right) 
\: \sigma_i \vrel}},
\label{eq:avratevrel}
\ee
where the we have defined the \emph{relative} velocity 
distribution:
\be
P_{\bm{x}, \mathrm{rel}}\left(\bm{v}_\mathrm{rel}\right)\equiv
\int{P_{\bm{x}, pair}\left(\bm{v}_\mathrm{cm},\bm{v}_\mathrm{rel}\right)
\diff^3 \bm{v}_\mathrm{cm} }.
\ee
Furthermore, Lorentz invariance requires that the annihilation does not
depend on the orientation of the relative velocity. Hence, to obtain
the average annihilation rate we just need to convolve 
the cross-section with the distribution of the magnitude of the relative
velocity.

\section{Density profile of the dark matter halo}
\label{sec:dmprofile}
The possibility to detect dark matter particles in the halo was initially
studied assuming that their distribution was that
of the Standard Halo~\cite{Drukier:1986tm}, which models the galaxy after
a singular isothermal sphere profile:
\be
\rho_{SIS} (r) = \frac{\sigma^2}{2 \pi G r^2}.
\label{eq:rhosis}
\ee
The increasing resolution of numerical N-body simulations in recent
years, however, has resulted in different density profiles
that provide a better description of the dark matter halo (see 
e.g.~\cite{Diemand:2009bm,Frenk:2012ph,Kuhlen:2012ft} for recent reviews). 
While less
singular than~\eq{rhosis}, the predictions from numerical experiments
suggest cuspy densities rising as $\sim 1/r$ towards the center.  
Several astrophysical observations, however, have shown that distributions
with a constant density core might provide a more accurate description of 
smaller halos, such as satellite galaxies of the Milky Way, or of the 
galactic center (see e.g.~\cite{deBlok:2009sp} for a review of observations
of LSB and gas-rich galaxies). Hence, we will consider both cored and cusped 
halos to model the spatial distribution of the dark matter particles. 

Although there is evidence from numerical simulations for the existence
of substructure in the form of unmixed dark matter clumps, we are primarily
concerned with annihilations from the smooth halo, which we take to be
spherically symmetric. As we argue in the discussion section, these
approximations provide a conservative estimate of the dark matter annihilation
rate.

We find it advantageous to work with dimensionless densities and 
potentials. 
Dehnen profiles~\cite{Dehnen:1993uh}, falling as $1/r^4$ at large distances, 
have a well defined total mass. This is not the case, however,
for the NFW~\cite{Navarro:1996gj} profile, whose mass grows logarithmically
at large distances. One then usually defines a virial radius $\rvir$
such that the average density contained in the spherical volume
$4/3 \pi \rvir^3$ is $\delta_c \approx 200$ times the critical density of the
universe, $\rho_\mathrm{crit}$. We will use the mass $\mvir$ contained 
in this region to normalize gravitational potentials and densities,
and we measure lengths in units of $\rvir$:
\be
x = \frac{r}{\rvir}.
\label{eq:length}
\ee

We consider spherically symmetric DM density profiles of the form
\be
\rho = \rho_0 \rhot(x,c,\cdots),
\label{eq:rhotilde0}
\ee
where $\rhot$ is dimensionless. Here, $c\equiv\rvir/a$ 
is the {\em concentration parameter}, with $a$ being a typical
length scale of the halo. For example, in two-power density models $a$ 
is the intermediate radius that marks the smooth transition between
the inner (possibly cuspy) power-law and the $1/r^{3-4}$ behaviour at 
large radii. For cored profiles, $a$ could represent the size of a central
finite density region, and there might be additional parameters in more
complicated multi-scale models, which we will omit from here on.

For a given profile,
\be
\mvir = 4 \pi \rvir^3 \rho_0 
\underbrace{\int_0^1{ x^2 \rhot(x,c) \diff x}}_{\equiv g(c)},
\ee
and we can rewrite eq.~(\ref{eq:rhotilde0}) as:
\be
\rho = \frac{\mvir}{4 \pi \rvir^3} \frac{1}{g(c)} \times \rhot(x,c).
\label{eq:rhotilde}
\ee
We will make frequent use of dimensionless analogues of 
the relative potential $\psi \equiv -\Phi$, 
where $\Phi$ is the gravitational potential per unit mass, and the relative 
energy $\epsilon=\psi - v^2/2$: 
\begin{align}
	\epst & \equiv \frac{\rvir}{G \mvir} \epsilon \nn
	\psit & \equiv \frac{\rvir}{G \mvir} \psi,
	\label{eq:epst}
\end{align}
where $G$ is Newton's gravitational constant.

The escape velocity at any given point of the halo is given by,
\be
v^\mathrm{max}(x)=\sqrt{2 \psit (x) \frac{G \mvir}{\rvir}}.
\ee
Since the potential, $\psit$, attains its maximum at $x=0$, the escape
velocity will be largest at the center of the halo.

The particular shape of the density profile can be inferred from observations
of stars tracing the gravitational potential of the halo or from numerical
N-body simulations. Although simulations favor a universal cusped profile, some
sub-galactic sized objects are better described by assuming the presence
of a central core. We, hence, consider both cusped and cored distributions.

\subsection{NFW profile}

The Navarro, Frenk, and White (NFW) profile,
\begin{align}
	\rho_\mathrm{NFW} &= \frac{\rho_0}{\frac{r}{a} \left(1+\frac{r}{a}
\right)^2} \nn
	&= \frac{\mvir}{4 \pi \rvir^3} \frac{1}{g_\mathrm{NFW}(c)} \times
\underbrace{\frac{1}{cx (1+c x)^2}}_{\equiv \rhot_\mathrm{NFW}},
\label{eq:rhonfw}
\end{align}
with 
\be
g_\mathrm{NFW}(c)= \frac{\log(1+c)-\frac{c}{1+c}}{c^3},
\ee
provides a good fit to DM-only N-body simulations over a wide range
of halo masses~\cite{Navarro:1996gj}. The associated gravitational
potential is
\be
\psit_\mathrm{NFW} = \frac{\log(1+c x)}{c^3 g_\mathrm{NFW}(c) x}.
\label{eq:psinfw}
\ee
With $c=10$ and a scale radius of $a=20$ kpc, the mass enclosed within
$\rvir=200$kpc is $\mvir=9.9\times 10^{11}M_\odot$. This falls within
the broad range of values, $5\times 10^{11}M_\odot \lesssim
\mvir \lesssim 3 \times 10^{12}M_\odot$, that have been inferred for the
Milky Way. Even though several studies have recently suggested
that the Milky Way might be less massive and more concentrated than 
previously thought (see e.g.~\cite{Deason:2012ky}), the values above are
adequate for our purposes.

To model a dwarf Spheroidal (dSph) satellite of the Milky Way 
with an NFW profile, we note that observations are consistent with
the known satellites having a mass of about 
$10^7 M_\odot$ within their central
$300$ pc~\cite{Strigari:2008ib}, while a scale radius $a=0.62$ kpc fits 
the observed radial velocity dispersion of the stars~\cite{Evans:2003sc}. 
The total mass of the dSph dark matter halos is difficult to determine, since
their extent beyond the observed stellar distributions is largely 
unknown~\cite{Walker:2012td}. For definiteness, 
we will assume that these objects extend up to $\rvir =3$ kpc, and contain
a mass of $\mvir=1.3 \times 10^8 M_\odot$, which is the value obtained by
extrapolating the NFW halo with the parameters determined above from more
robust observations of the central region. Hence, $c=4.8$ in~\eq{rhonfw},
which is substantially lower than the typical values $c\sim 20$ for halos
of this size found in simulations~\cite{2008MNRAS.391.1940M}. 

Since
\be
\psit^\mathrm{max}=\psit (0) = \frac{1}{c^2  g_\mathrm{NFW}(c) },
\ee
the maximum velocity of any bound particle in an NFW halo is
\be
v^\mathrm{max}_\mathrm{NFW}=\sqrt{\frac{2}{c^2  g_\mathrm{NFW}(c) } 
\frac{G \mvir}{\rvir}}.
\label{eq:vmaxnfw}
\ee
For the Galaxy, $v^\mathrm{max}_\mathrm{NFW} \approx 537.3\: \mathrm{km/s}$, and
for a typical dwarf, $v^\mathrm{max}_\mathrm{NFW}\approx 43.7\; \mathrm{km/s}$.

\subsection{Einasto profile}
In addition to describing the luminosity profiles of early-type galaxies
and bulges and the surface density of hot gas in clusters, the Einasto
profile is as good a fit as the NFW profile, if not better, to simulated
galaxy-sized dark matter halos:
\begin{align}
	\rho_\mathrm{Ein} &= \rho_0 \exp\left(-\frac{2}{\gamma} \left[ \left(
\frac{r}{a}\right)^\gamma -1 \right] \right) \nn
	&= \frac{\mvir}{4 \pi \rvir^3} \frac{1}{g_\mathrm{Ein}(c)} \times
\underbrace{\exp\left(-\frac{2}{\gamma} \left[ \left(
c x \right)^\gamma -1 \right] \right) }_{\equiv \rhot_\mathrm{Ein}}.
\label{eq:rhoein}
\end{align}
Here,
\be
g_\mathrm{Ein}(c)= \frac{\left(\frac{2}{\gamma}\right)^{-3/\gamma} 
\exp \left(2/\gamma\right)}{c^3 \gamma} \left(\Gamma\left(\frac{3}{\gamma}
	\right) - \Gamma\left(\frac{3}{\gamma}, \frac{2 c^\gamma}{\gamma}
\right)\right),
\ee
where $\Gamma(x)$ and $\Gamma(s,x)$ are the usual gamma function and
the incomplete gamma function respectively~\cite{Cardone:2005ne}. A
value of $\gamma=0.17$ provides a good fit to galactic- and cluster-sized
halos in N-body simulations~\cite{Navarro:2003ew,Graham:2005xx}. 
As with the NFW profile, we take $c=10$ and $\rvir \approx 200$ kpc for 
the Milky Way, which falls roughly along the concentration-mass relation
for the WMAP5 cosmology~\cite{Duffy:2008pz}.

The associated potential can be written as~\cite{Cardone:2005ne}:
\begin{align}
	\psit_\mathrm{Ein} = &\frac{c}{\Gamma\left(3/\gamma\right)
-\Gamma\left(3/\gamma, 2 c^\gamma/\gamma\right)} \nn  
& \times \left[  \frac{\Gamma\left(3/\gamma
	\right) - \Gamma\left(3/\gamma, 2 (c x)^\gamma/\gamma \right)}{c x} +
	\left(\frac{2}{\gamma}\right)^{1/\gamma}  
	\Gamma\left(2/\gamma, 2 (c x)^\gamma/\gamma\right)
 \right].
\label{eq:psiein}
\end{align}

\subsection{Burkert profile}

The fact that the best NFW fit to the observations of dSphs can
be far less concentrated than expected from simulations suggests that
the NFW profile provides a poor fit to the dynamics of some
dSphs. In particular, detections of distinct stellar sub-populations 
provide mass estimates at different radii for Fornax and Sculptor
that are consistent with cored potentials, but largely incompatible with
cusped profiles~\cite{Walker:2011zu,Amorisco:2012rd}.

The Burkert profile is a cored profile that appears to provide a good fit 
to the DM distribution
in dSph galaxies~\cite{Burkert:1995yz}. Its density is given by
\begin{align}
	\rho_{\mathrm{Bur}} &= \frac{\rho_0}{\left(1+\frac{r}{a}\right)
\left(1+\frac{r^2}{a^2}\right)} \nn
&= \frac{\mvir}{4 \pi \rvir^3} \frac{1}{g_\mathrm{Bur}(c)} \times
\underbrace{\frac{1}{(1+cx) \left(1+(c x)^2\right)}}_{\equiv 
\rhot_\mathrm{Bur}},
\label{eq:rhobur}
\end{align}
with 
\be
g_\mathrm{Bur}(c)= \frac{\log(1+c^2)+ 2 \log(1+c) - 2\arctan (c)}{4 c^3}.
\ee

The gravitational potential generated by this profile is
\begin{align}
	\psit_\mathrm{Bur} = \frac{\pi c x - 2 (1+cx) \arctan(cx)+2(1+cx)
	\log(1+cx)+(1-cx)\log \left(1+(cx)^2\right)}{4 c^3 x
	g_\mathrm{Bur}(c)}.
\label{eq:psibur}
\end{align}

A scale radius of $a=650\,\mathrm{pc}$ and $\rho_0=1.8\times 10^8 \,M_\odot$
in~\eq{rhobur} were found in~\cite{Salucci:2011ee} to fit
the kinematics of Draco. With these parameters, 
and for $\rvir=3\,\mathrm{kpc}$, we obtain $\mvir=6 \times 10^8 \,M_\odot$, which
is slightly larger than our NFW model of a dwarf. To make comparisons
easier, we keep $c=4.6$, but we rescale down $\rho_0$ to match
$\mvir=1.3 \times 10^8 \,M_\odot$.

\subsection{Galactic bulge and disk}
In addition to the dark halo, the gravitational potential of the Milky Way
receives contributions from stars in the disk and the bulge. Indeed, the
baryonic contribution to the gravitational field is the dominant one in
the central region of the Galaxy and will play a crucial role in
determining the dark matter velocity distribution.
We model the baryons by adding
a central bulge and a disk. Following~\cite{Strigari:2009zb}, we take a
spherically-symmetric Hernquist potential for the bulge,
\be
\psit_\mathrm{bulge}=\frac{M_\mathrm{bulge}}{\mvir} \frac{1}{x + \frac{c_0}{
\rvir}},
\label{eq:psibulge}
\ee
where $c_0 \sim 0.6$kpc and $M_\mathrm{bulge}=1.5\times 10^{10}M_\odot$.
We represent the disk by a spherical distribution that approximates the
mass and circular velocity of the exponential disk:
\be
\psit_\mathrm{disk}=\frac{M_\mathrm{disk}}{\mvir} 
\frac{1-\exp\left(-\frac{\rvir \: x}{b_\mathrm{disk}}\right)}{x},
\label{eq:psidisk}
\ee
with $b_\mathrm{disk} \sim 4$ kpc and $M_\mathrm{disk}=5\times 10^{10}\,M_\odot$.
Although the galactic disk is certainly flattened, the distribution
in~\eq{psidisk} contains the same amount of mass interior to $x$ as an
exponential disk and matches its circular speed with error no more than $\sim 15\%$ 
(c.f. fig. 2.17 in \cite{binney}). We discuss the effects of a flattened
potential in section~\ref{sec:discussion}.

\section{Phase-space distribution function}
\label{sec:vdistribution}
As explained in section~\ref{sec:dmflux}, the kinematics of the
annihilating dark matter particles in the gravitational field of the halo
can be fully described by the DF, $f$. 

Most studies of DM indirect detection implicitly take $f$ to be the 
product of one of the density profiles discussed in Sec.~\ref{sec:dmprofile}
and a Maxwell-Boltzmann velocity distribution independent of the position
in the halo. As discussed below, such a distribution function
cannot in general describe a system of collisionless particles moving 
under the influence of a smooth gravitational potential. Indeed, only
for the density in eq.~(\ref{eq:rhosis}) does a Maxwell-Boltzmann velocity distribution
satisfy the Boltzmann equation.
The results of numerical simulations and better observations have,
for the most part, been incorporated in the spatial part of $f$, without
updating the velocity distribution. For the most discussed
neutralino models featuring a mostly s-wave velocity independent
cross-section one can ignore this discussion, since the production
rate in eq.~(\ref{eq:avratevrel}) does not depend on the shape
of the velocity distribution as long as it is properly normalized.
This is no longer the case for the DM scenarios 
with Sommerfeld enhancement that have been recently discussed in connection
with the observed cosmic-ray anomalies. A more practical reason for neglecting
the velocity distribution is that it is much more difficult to measure than
the density. The resolution of numerical simulations has increased
dramatically in recent years, but the number of particles is still too
small to sample the possible velocities at each point in the synthetic halo.
To estimate the velocity distribution at the particular location of the
solar system, an average over $100$ randomly distributed
sample spheres centered at $8.5$ kpc was performed in~\cite{Kuhlen:2009vh}
to capture about $10^4$ particles among the billion particles in
{\em Via Lactea II}, an N-body simulation of a Milky-Way-size galaxy. 
These results provide insight into the non-gaussian
shape of the velocity distribution and on the influence of substructure,
but an extension to map the inner and outer parts of the halo is 
clearly unfeasible. An empirical fit to the local distribution was
found in~\cite{Mao:2012hf} from a suite of cosmological simulations,
and a full phase-space distribution consistent with a fit to the
observed circular rotation of the Galaxy was considered 
in~\cite{Chaudhury:2010hj}. 

Such studies of the local distribution
are motivated by the conflicting results of direct detection experiments,
which are sensitive to the velocity of the DM particles even for 
s-wave annihilating DM. We here extend the study of the velocity distribution
to the rest of the halo, as required in studies of DM indirect detection, 
by using the formalism developed by Eddington. Although, we leave 
the consideration of a velocity anisotropy 
or a fit to numerical simulations for future work~\cite{fhprogress}, our
method captures the effects of deviations from gaussianity found in studies
of the local neighborhood, while maintaining a self-consistent phase-space
distribution that satisfies the Boltzmann equation.
   
For a spherical system confined by a known gravitational potential $\psi$,
we can find a unique ergodic (i.e. isotropic in velocity space) 
distribution function using Eddington's result~\cite{binney}. In this case,
$f$ is a function of the energy per unit mass and can be written as
\be
	f\left(\epst \right) = \frac{1}{8 \sqrt{2} \pi^3 \sqrt{G^3 \rvir^3
	\mvir}} \frac{1}{g(c)} \times \tilde{f} \left(\epst\right),
\ee
where 
\be
\tilde{f} \left( \epst  \right) \equiv 
\int_0^{\epst}{ \frac{\diff \psit}{\sqrt{\epst-\psit}} 
\frac{\diff^2 \rhot}{\diff \psit^2}}.
\label{eq:eddingtonp}
\ee

\begin{figure}[htb]
	\begin{center}
		\includegraphics[width=0.8\textwidth]{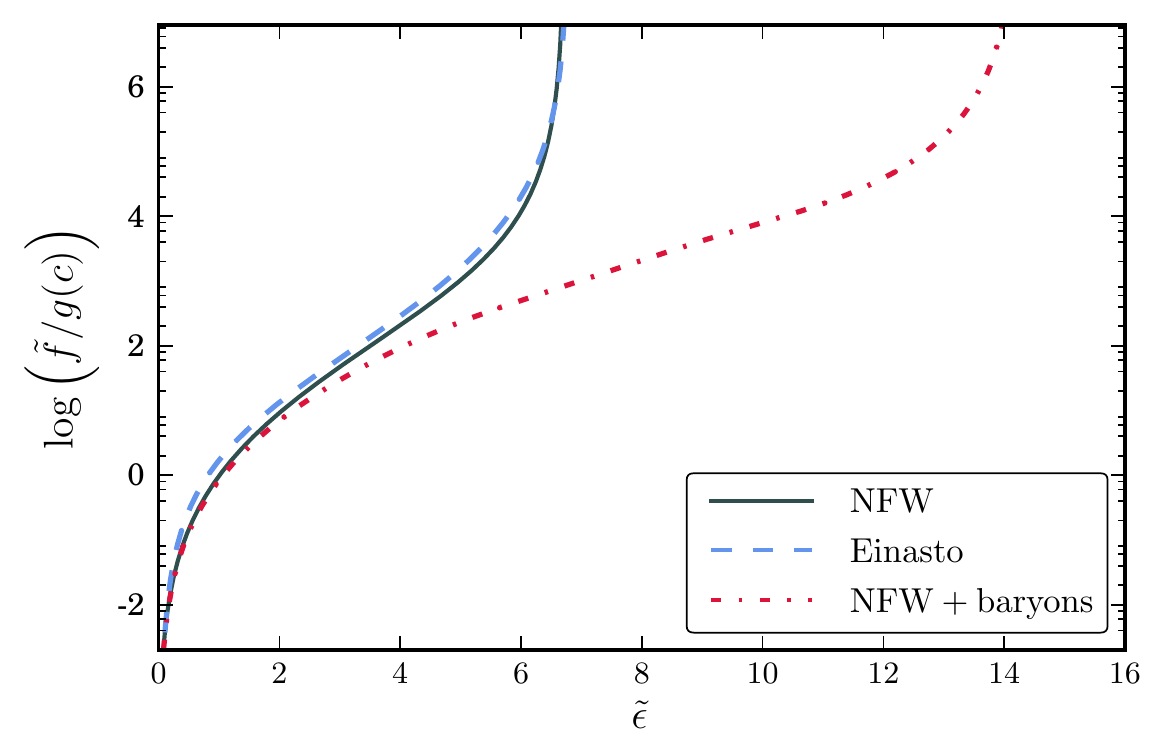}
	\end{center}
	\caption{Phase-space distribution function of the dark matter
		in the Galaxy, \eq{eddington}, 
		assuming an NFW profile, with and without a baryonic
	disk, or a dark matter only Einasto profile.}
	\label{fig:dfgal}
\end{figure}

\begin{figure}[htb]
	\begin{center}
		\includegraphics[width=0.8\textwidth]{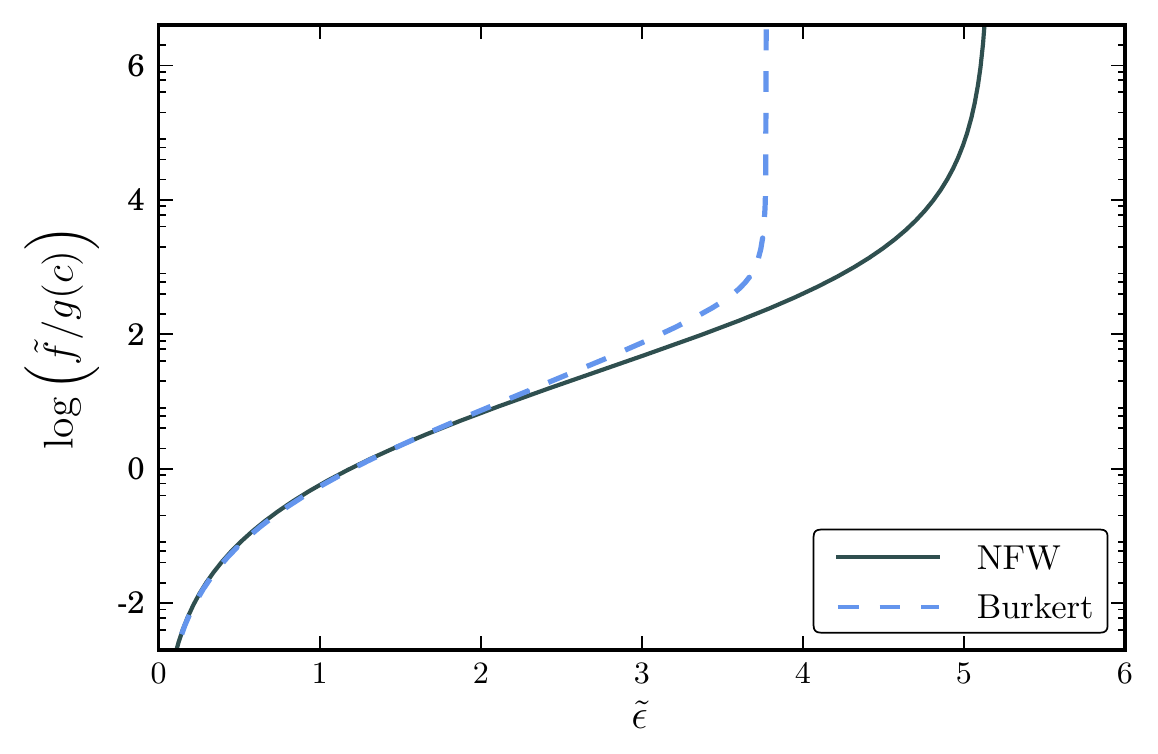}
	\end{center}
	\caption{Phase-space distribution function of the dark matter
		in a dSph with an NFW or a cored Burkert profile.}
	\label{fig:dfdwarf}
\end{figure}

It is not possible in general to invert $\psit(x)$ in order to express
$\rhot$ as a function of $\psit$. We can, nevertheless, express the
derivative in \eq{eddingtonp} as
\be
\frac{\diff^2 \rhot}{\diff \psit^2} = \left(\frac{\diff \psit}{\diff x}
\right)^{-2} \left(\frac{\diff^2 \rhot}{\diff x^2}-
	\left(\frac{\diff \psit}{\diff x}
\right)^{-1} \frac{\diff^2 \psit}{
\diff x^2} \frac{\diff \rhot}{\diff x}\right).
\ee
Then~\eq{eddingtonp} reads:
\be
\tilde{f} \left( \epst  \right) = 
\int_{0}^{\epst}{ 
		\frac{\diff \psit}{\sqrt{\epst-\psit}} 
	\left(\frac{\diff \psit}{\diff x'}
	\right)^{-2} \left(\frac{\diff^2 \rhot}{\diff x'{}^2}-
	\left(\frac{\diff \psit}{\diff x}\right)^{-1} 
		\frac{\diff^2 \psit}{
\diff x'{}^2} \frac{\diff \rhot}{\diff x'}\right)},
\label{eq:eddington}
\ee
where the integrand and the limits of integration are 
a function of the position $x$. We numerically solve $\psit = \psit(x)$ for
$x$ when performing the integration in \eq{eddington} to find $\tilde{f}
\left( \epst \right)$. We show in figs.~\ref{fig:dfgal}~and~\ref{fig:dfdwarf}
the distribution function for the Galaxy and for a classical dSph.

The three-dimensional velocity distribution can be written as
\begin{align}
	P_{\bm{x}} (\bm{v}) \diff^3 \bm{v}& =
	\frac{1}{\sqrt{8}\pi^2} \sqrt{\left(\frac{\rvir}{G \mvir}\right)^{3}}
	\frac{\tilde{f}\left(\psi - v^2/2\right)}{\rhot(\bm{x})} 
	\diff^3 \bm{v}\nn
&=\frac{1 }{\sqrt{8}\pi^2} \frac{\tilde{f}\left(\psit 
- \tilde{v}^2/2\right)}{\rhot(\bm{x})} 
\diff^3 \bm{\tilde{v}}, 
	\label{eq:fv3d}
\end{align}
which is a function of $v^2$ for an isotropic spherical system, 
$P_{\bm{x}}(\bm{v}) = P_{\bm{x}} \left(v^2\right)$, and
$\tilde{v}{}^2\equiv \frac{\rvir}{G \mvir} v^2$. 

%{\bf FF: This is not actually the 1D distribution function, but the
%distribution of $v$, but is the one to be compared to MB. 
%The 1D df would be the distribution in one of the components.}
%{\bf FF: Get rid of the tilde distribution functions, which are only used
%for calculation}.
One can obtain the one-dimensional distribution of the magnitude of
the velocity by performing the angular integration in $\bm{v}$-space:
\begin{align}
	P_{r} \left(v^2\right) \diff v &= \frac{\sqrt{2}}{\pi} 
	\sqrt{\left(\frac{\rvir}{G \mvir}\right)^{3}}\: v^2 
\frac{\tilde{f}}{\rhot}\diff v \nn
&=\frac{\sqrt{2}}{\pi} 
\tilde{v}{}^2 
\frac{\tilde{f}}{\rhot}\diff \tilde{v}.
\label{eq:fv1d}
\end{align}
This distribution function is normalized as follows:
\be
\int_0^\infty{P_{r} \left(v^2\right) \diff v }= 1,
\ee
where the upper integration limit can be replaced by the maximum velocity,
$\sqrt{2 \psi(r)}$. 

\begin{figure}[htb]
	\begin{center}
		\includegraphics[width=0.8\textwidth]{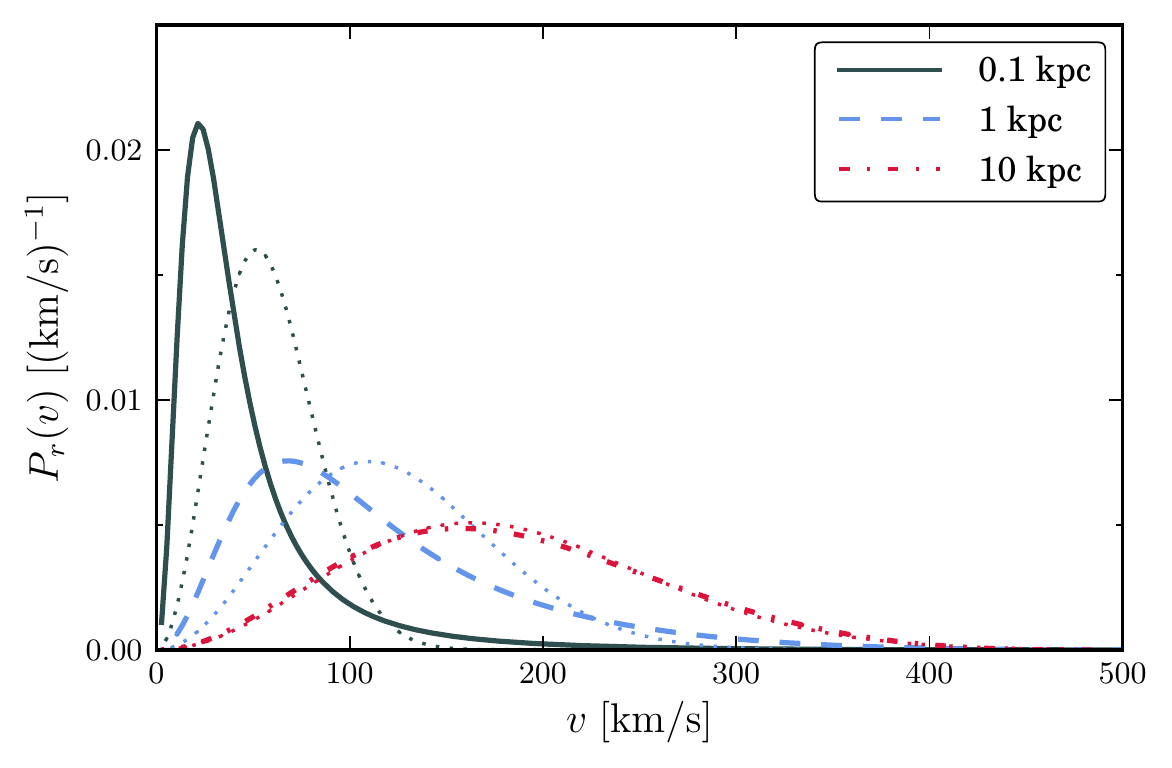}
	\end{center}
	\caption{Velocity distribution, \eq{fv1d}, for the an NFW galaxy 
		without baryons. The dotted lines show a Maxwell-Boltzmann 
		distribution with the same velocity dispersion.}
	\label{fig:fv_nfwgal}
\end{figure}

\begin{figure}[htb]
	\begin{center}
		\includegraphics[width=0.8\textwidth]{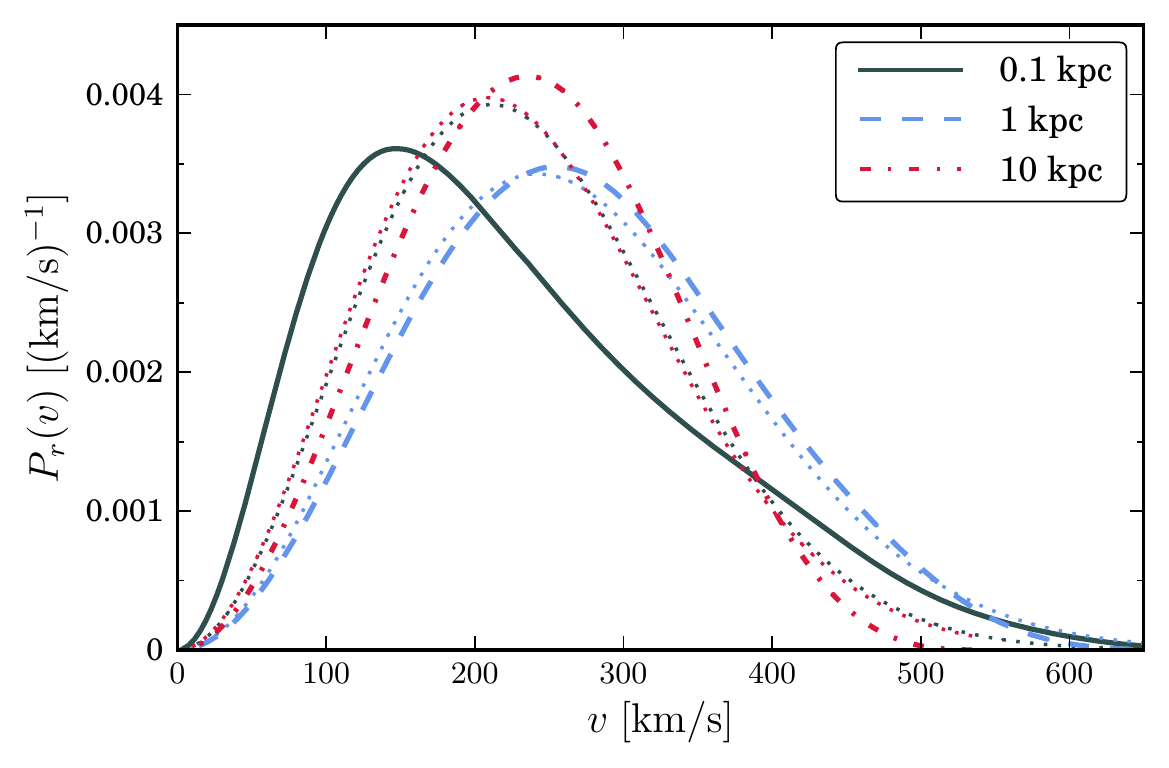}
	\end{center}
	\caption{Velocity distribution, \eq{fv1d}, for an NFW galaxy 
	with disk and bulge~eqs.~(\ref{eq:psidisk},\ref{eq:psibulge}).}
	\label{fig:fv_nfwgaldb}
\end{figure}

\begin{figure}[htb]
	\begin{center}
		\includegraphics[width=0.8\textwidth]{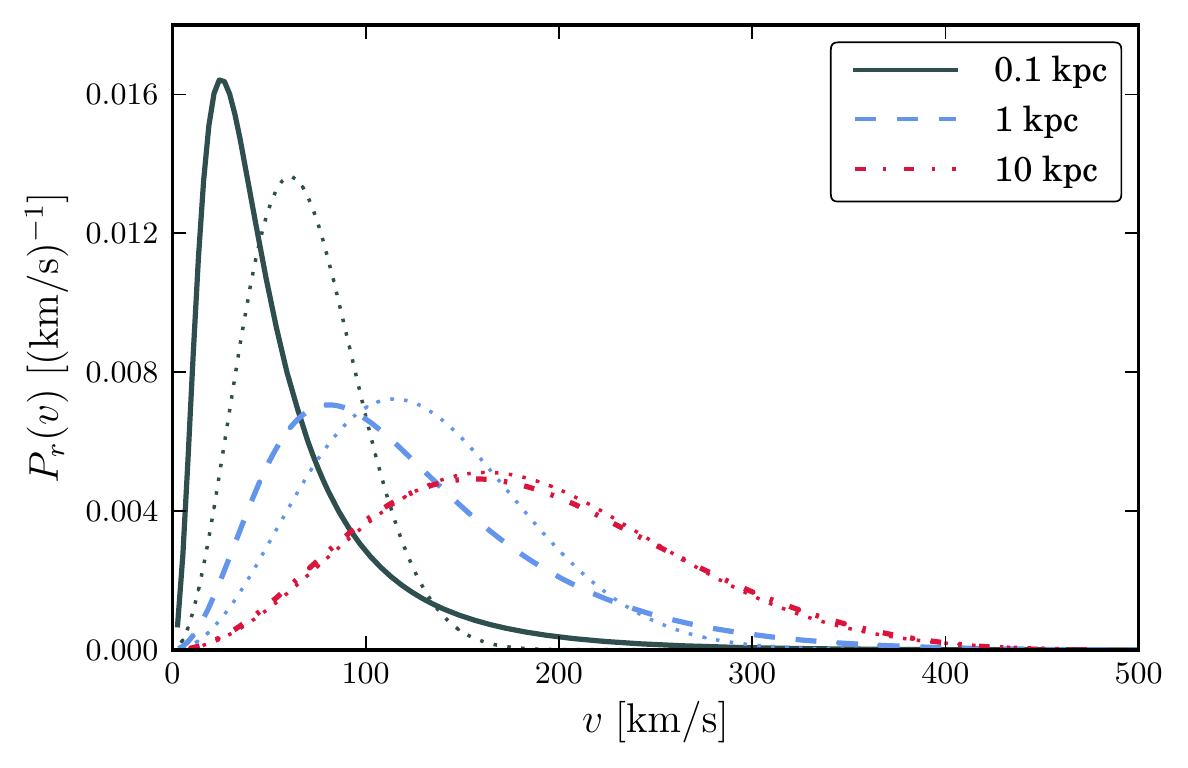}
	\end{center}
	\caption{Velocity distribution, \eq{fv1d}, for an Einasto galaxy-sized
	dark matter halo.}
	\label{fig:fv_eingal}
\end{figure}

\begin{figure}[htb]
	\begin{center}
		\includegraphics[width=0.8\textwidth]{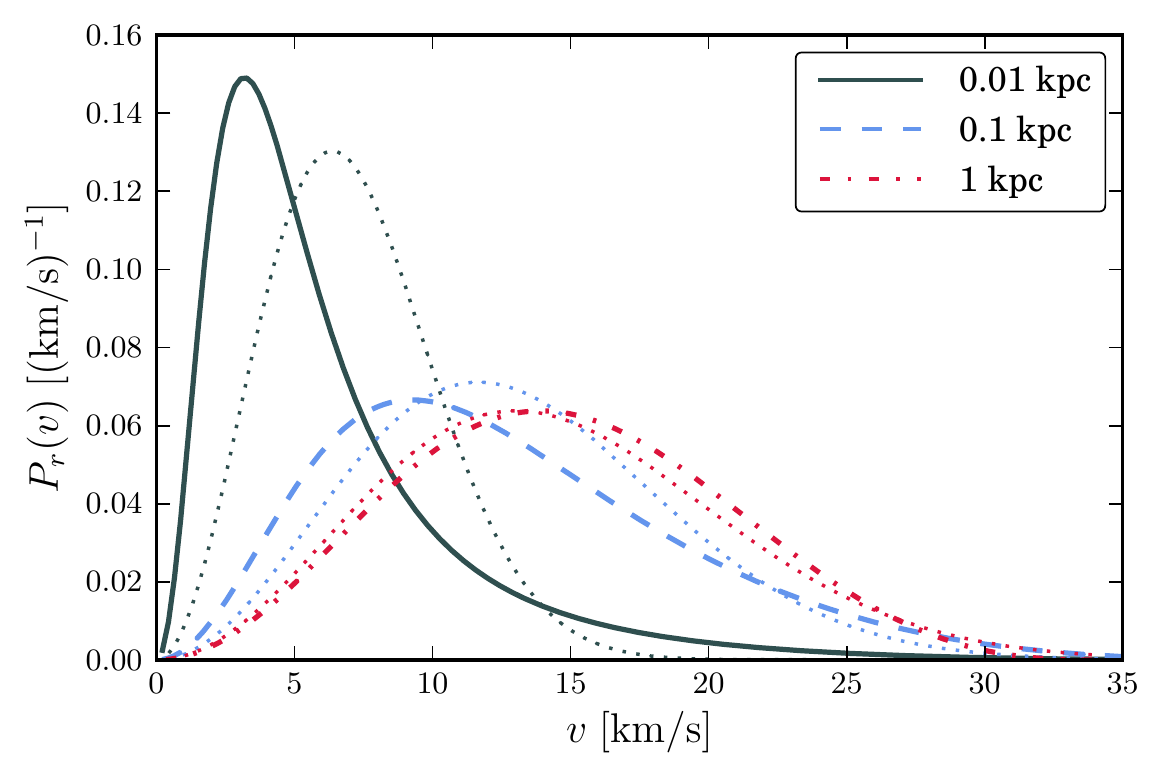}
	\end{center}
	\caption{Velocity distribution, \eq{fv1d}, for a classical dSph
	satellite of the Milky Way with an NFW profile.}
	\label{fig:fv_nfwdwarf}
\end{figure}

\begin{figure}[htb]
	\begin{center}
		\includegraphics[width=0.8\textwidth]{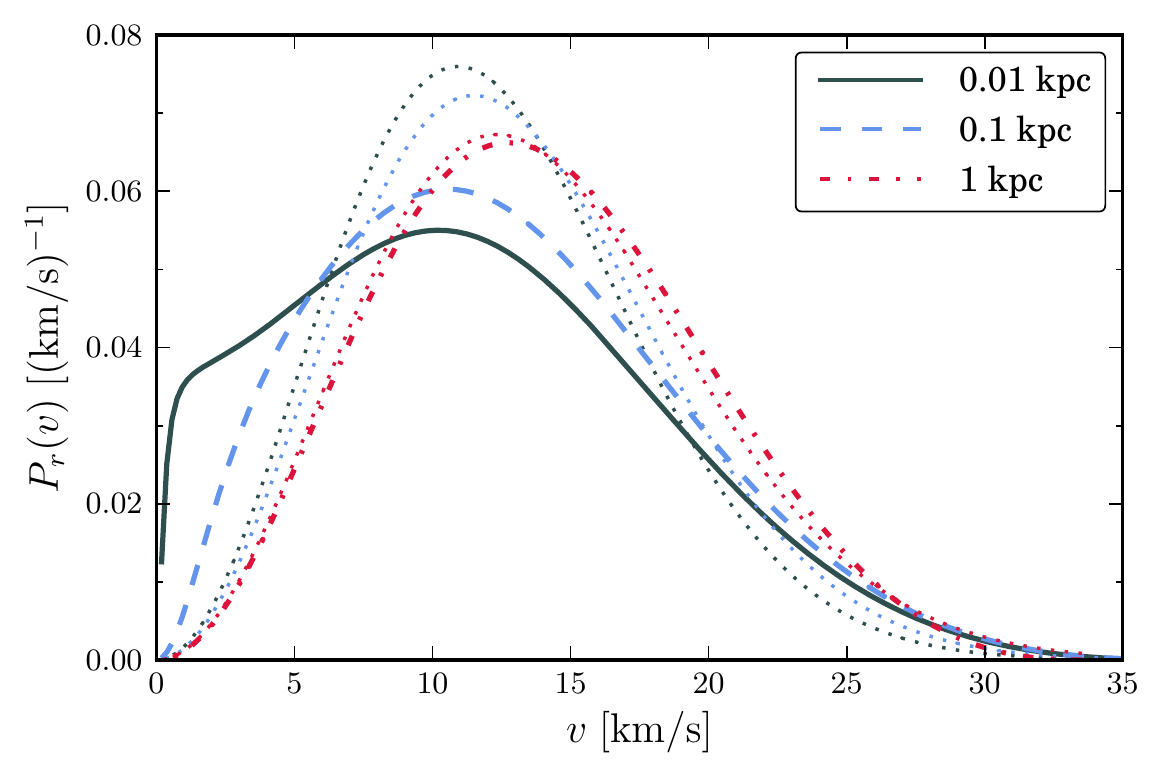}
	\end{center}
	\caption{Velocity distribution, \eq{fv1d}, for a classical dSph
	satellite of the Milky Way with a more appropriate Burkert profile.}
	\label{fig:fv_burkert}
\end{figure}

Figs.~\ref{fig:fv_nfwgal}, \ref{fig:fv_nfwgaldb},
\ref{fig:fv_eingal},~\ref{fig:fv_nfwdwarf},~and~\ref{fig:fv_burkert}
show the distribution function in~\eq{fv1d}. At large distances, the 
distribution can be well-approximated by a Maxwell-Boltzmann shape. However,
as we move towards the center of the halo, large departures from gaussianity
are evident. In addition, as we move closer to the center of the halo,
the actual distribution function shifts to
lower values of the velocity, in stark contrast
with the uniform Maxwell-Boltzmann (MB) shape with constant velocity 
dispersion,
which is usually assumed in studies of indirect detection of dark matter.

In the Galaxy, the added gravitational potential of the disk and the bulge
contributes to alleviate the cooling of dark matter particles towards the
center. Although this improves the agreement with a description based on
a constant MB distribution at moderate distances, large departures from
gaussianity are still evident within the central $1$ kpc.

\subsection{The relative velocity distribution}

The yield of SM particles from dark matter annihilation depends on the
distribution of {\em relative} velocities at a given location, as
shown in section~\ref{sec:dmflux}.
For a spherical systems with an ergodic
distribution function, $\tilde{f}= \tilde{f}\left(v^2\right)$ at any point
in the halo. Then, the individual velocity distributions in~\eq{vpair}
do not depend on the six components of the velocities, but only on
the combinations:
\be
P_{\bm{x}}\left(v_{1,2}^2\right) = P_{\bm{x}}\left(v_{\mathrm{cm}}^2 +
	v_{\mathrm{rel}}^2/4 \pm \bm{v}_{\mathrm{cm}} \cdot 
	\bm{v}_{\mathrm{rel}} \right).
\ee
Using spherical coordinates in $\bm{v}_\mathrm{rel}$-space, with
the $z$-axis in the direction of the relative velocity vector,
\begin{align}
	P_{\bm{x}, \mathrm{rel}}\left(v_\mathrm{rel}\right) \equiv 
8 \pi^2  v_\mathrm{rel}^2 \int_0^{\infty}{\diff v_{\mathrm{cm}} \,
v_{\mathrm{cm}}^2 \int_{-1}^{1} {\diff z }}\,& 
P_{\bm{x}}\left( v_{\mathrm{cm}}^2 +
v_{\mathrm{rel}}^2/4 + v_{\mathrm{cm}}v_{\mathrm{rel}} z \right) \times \nn 
&	P_{\bm{x}}\left( v_{\mathrm{cm}}^2 +
v_{\mathrm{rel}}^2/4 - v_{\mathrm{cm}}v_{\mathrm{rel}} z \right), 
\label{eq:fvrelp}
\end{align}
where $z$ is the cosine of the angle between the relative and the center
of mass velocities. In appendix~\ref{sec:reldf} we write the expression
above explicitly in terms of the individual phase-space distribution 
function.

\paragraph{Results for a Maxwell-Boltzmann distribution}

We can use the formalism developed above to recover the well-known results
of the Standard Halo~\cite{Drukier:1986tm}, which models the galaxy after
a singular isothermal sphere profile.

The phase-space distribution function that gives rise to the density in
\eq{rhosis} is~\cite{binney}
\be
f(\epsilon)=\frac{\rho_1}{\left(2 \pi \sigma^2\right)^{3/2}} 
\exp\left(\frac{\psi-\frac{v^2}{2}}{\sigma^2}\right),
\ee
which results in a velocity distribution of the Maxwell-Boltzmann type:
\be
P_{\bm{x}}^{MB} (\bm{v})= \frac{1}{\left(2 \pi \sigma^2\right)^{3/2}} 
\exp\left(\frac{-v^2}{2 \sigma^2}\right).
\label{eq:fvmb}
\ee
Note that the MB velocity distribution does not depend on the spatial
coordinates, i.e. $\sigma$ is a constant.

A straightforward application of \eq{fvrelp} gives the relative velocity
distribution:
\be
P_{\mathrm{rel}}^{MB} \left(v_{\mathrm{rel}} \right) =
		4 \pi v_\mathrm{rel}^2 
		\frac{1}{\left(2 \pi \left(2\sigma^2\right)\right)^{3/2}} 
		\exp\left(\frac{-v_{\mathrm{rel}}^2}{2 
			\left(2\sigma^2\right)}
			\right).
\label{eq:fvrelmb}
\ee
It is a particularity of the MB distribution that the relative velocity
distribution also has the same functional form, i.e. it is another
MB distribution. The one-dimensional relative velocity dispersion is,
of course, doubled in the center of mass with respect to that of the 
individual particles. 

The velocity dispersion $\sigma$ can be related
to observable quantities, such as the mass interior to radius $r$ or
the circular speed. Alternatively, it can be determined from N-body
simulations, as was done in~\cite{Kuhlen:2009kx}, where a constant
MB distribution was used to estimate the Sommerfeld correction to
{\em Via Lactea II}, an N-body simulation of a Milky-Way-size galaxy.

\paragraph{Relative velocity distribution for an NFW or Burkert profile}

\begin{figure}[htb]
	\begin{center}
		\includegraphics[width=0.8\textwidth]{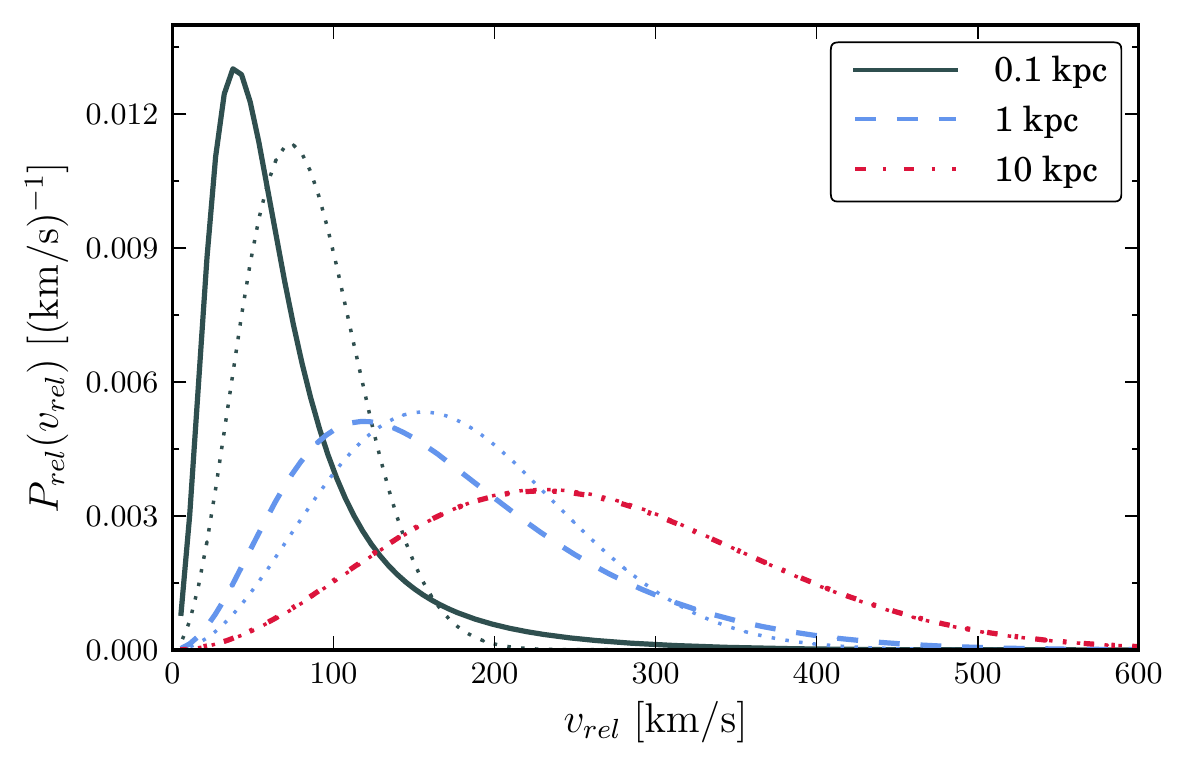}
	\end{center}
	\caption{Relative velocity distribution, \eq{fvrelp}, for an NFW
		dark matter halo only. The dotted lines show a 
		Maxwell-Boltzmann distribution with the same velocity 
	dispersion: $\sigma = 51.8 \mathrm{km/s}, 110.1 \mathrm{km/s},
163.2\mathrm{km/s}$ for $r=0.1 \mathrm{kpc},1 \mathrm{kpc}, 10 \mathrm{kpc}$. }
	\label{fig:frel_nfwgal}
\end{figure}

\begin{figure}[htb]
	\begin{center}
		\includegraphics[width=0.8\textwidth]{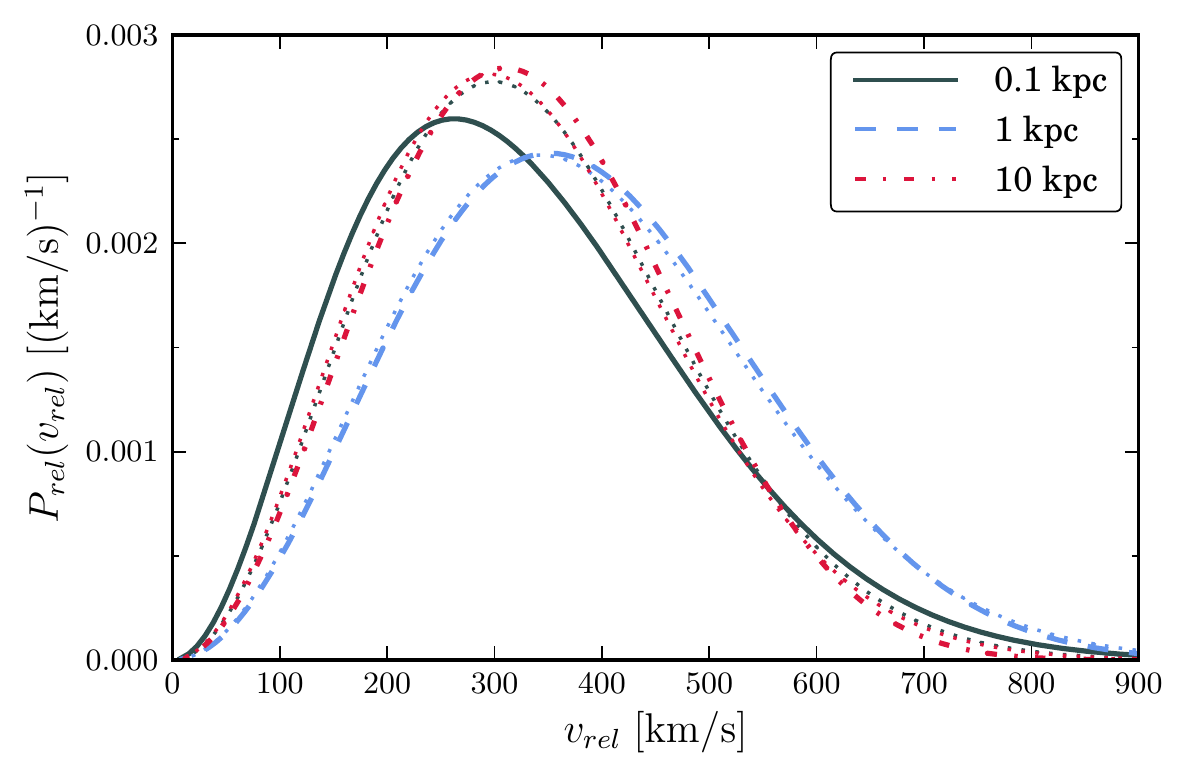}
	\end{center}
	\caption{Relative velocity distribution, \eq{fvrelp}, for an NFW Galaxy 
		with disk and bulge. The dotted lines show a 
		Maxwell-Boltzmann distribution with the same velocity 
	dispersion: $\sigma = 211.4 \mathrm{km/s}, 242.2 \mathrm{km/s},
208.8\mathrm{km/s}$ for $r=0.1 \mathrm{kpc},1 \mathrm{kpc}, 10 \mathrm{kpc}$. }
	\label{fig:frel_nfwgaldb}
\end{figure}

\begin{figure}[htb]
	\begin{center}
		\includegraphics[width=0.8\textwidth]{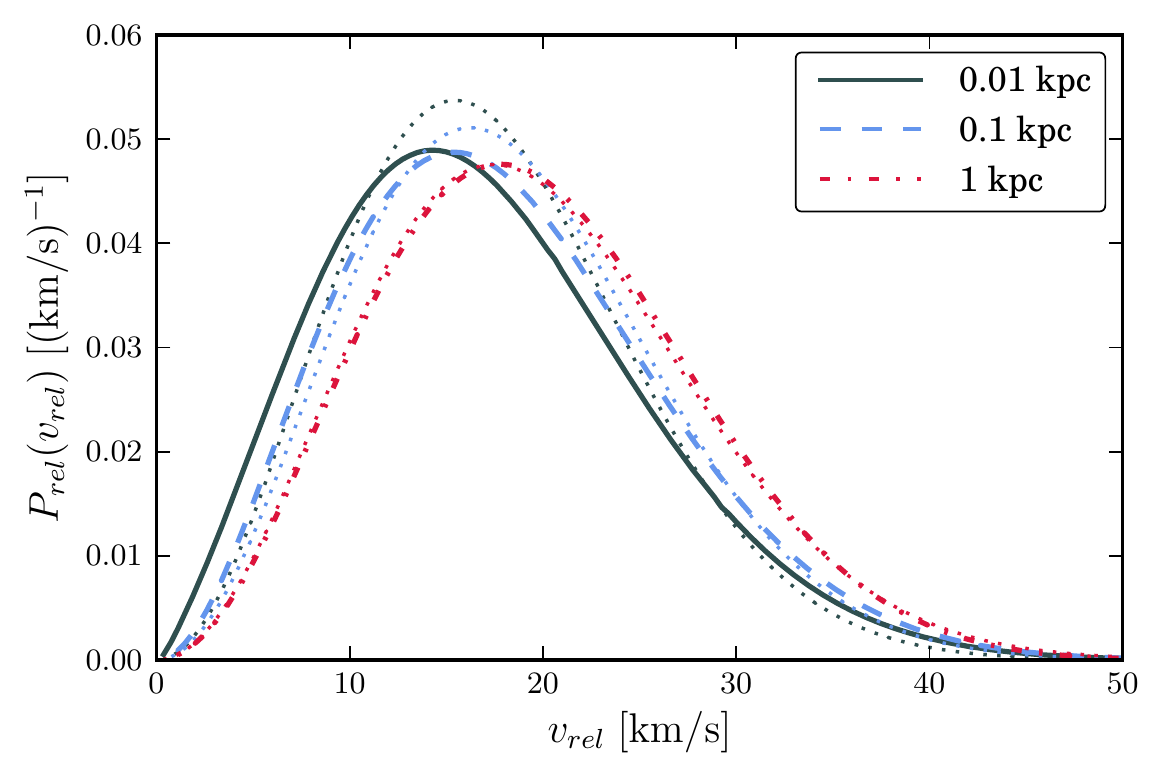}
	\end{center}
	\caption{Relative velocity distribution, \eq{fvrelp}, for a classical 
		dSph satellite of the Milky Way with a Burkert
		profile. The dotted lines show a 
		Maxwell-Boltzmann distribution with the same velocity 
	dispersion: $\sigma = 10.9 \mathrm{km/s}, 11.5 \mathrm{km/s},
12.3 \mathrm{km/s}$ for $r=0.01 \mathrm{kpc},0.1 \mathrm{kpc}, 
1 \mathrm{kpc}$. }
	\label{fig:frel_burkert}
\end{figure}

Figs.~\ref{fig:frel_nfwgal}, 
\ref{fig:frel_nfwgaldb},~and~\ref{fig:frel_burkert}
show the distribution function for the relative velocity in~\eq{fvrelp}.
As expected, the velocity
dispersion changes with the distance to the center of the halo,
and the distribution can be highly non-gaussian. 
For an NFW density profile, a Maxwell-Boltzmann distribution function
is a good approximation to the exact relative velocity distribution for
distances of the order of the scale radius ($\sim 20$ kpc for the Galaxy
and $\sim 0.15$ kpc for a dSph) and larger. This is reasonable, since the density 
profile behaves like $\propto 1/r^2$ in this region, similar to the SIS.

Adding a baryonic disk and bulge to the Galaxy 
increases the gravitational potential,
which raises the relative velocity dispersion. Also, the Maxwell-Boltzmann
distribution is a better fit for distances above $\sim 1$ kpc. However, as 
we enter the innermost region, the departures from gaussianity are evident
again.

\subsection{Consistency with Jeans analysis}
The distribution function contains all the information about the behaviour
of the DM particles in phase-space. A limited understanding of the dependence 
of the velocity distribution on the position of the halo can be obtained in 
a more direct way by taking velocity moments of the collisionless Boltzmann 
equation. In this manner, one obtains Jeans equation that can be used
to compute the one-dimensional velocity dispersion. 

For a spherical system with an isotropic velocity distribution, the Jeans
equation reads~\cite{binney}
\begin{align}
	\bar{v}_r{}^2  (r) &= \frac{1}{\rho(r)} \int_r^\infty{\diff r'
\frac{\diff \Phi}{\diff r'} \rho(r')} \nn
&= \frac{G \mvir}{\rvir} \frac{1}{\rhot (x)} \int_\infty^x{\diff x' 
\frac{\diff \psit}{\diff x'} \rhot(x')},
\label{eq:jeans}
\end{align}
which, for an NFW halo, can be expressed analytically in terms of 
elementary and polylogarithm funcions. 

On the other hand, we can find the dispersion directly from the distribution
function in~\eq{fv1d}:
\be
3 \bar{v}_r^2  \left(r\right) = \int_0^{\sqrt{2 \psit}}{\diff v\, v^2 P_r(v)}.
\label{eq:jeansnum}
\ee

To test our numerical procedure, we check that the results of \eq{jeans}
and~\eq{jeansnum}
%, shown in fig.~\ref{fig:jeans1}, 
agree within numerical accuracy.

\begin{figure}[htb]
	\begin{center}
		\includegraphics[width=0.8\textwidth]{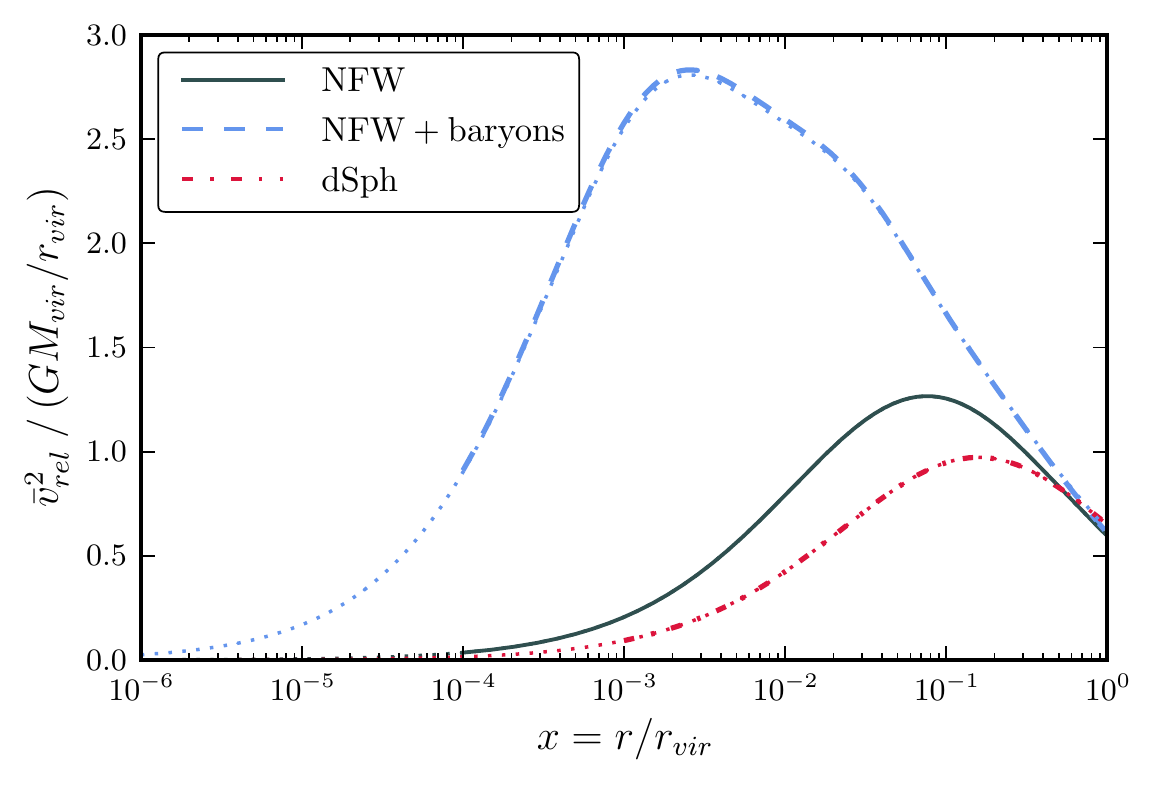}
	\end{center}
	\caption{One dimensional {\em relative} velocity dispersion. The
		thick lines show the result of using the exact velocity
		distribution, while the thinner dotted lines use
		a Jeans analysis and assume that
		the one-particle distribution is MB, like in \eq{fvrelmb}.
		Note that in physical units, the line corresponding
to the dSph would be down by a factor of $\mvir^{dSph} \rvir^{gal} / 
\mvir^{gal}\rvir^{dSph} \sim 1.5\times 10^3$}
	\label{fig:jeansrel}
\end{figure}

Since the one-particle distribution is not Maxwellian, we cannot in principle
use \eq{fvrelmb} to find the dispersion in the relative velocity. However, as
\fig{jeansrel} shows, the approximation $\bar{v}_{rel}^2 = 2 \bar{v}_1^2$ 
holds well for a wide range of distances. This behaviour is consistent
with Jeans type analyses to higher orders that show how the lowest order
moment hardly constrains the form of the potential~\cite{1994MNRAS.271..949M}.
As a consequence, for dark matter particles annihilating via a {\em p-wave}
process (or, more trivially, {\em s-wave}), we can get accurate fluxes
simply with a Jeans analysis. This was the strategy adopted 
in~\cite{Robertson:2009bh}, where it was also pointed out that 
non-gaussianities at small galactocentric distances cannot be captured
with this formalism. This is particularly important for dark matter particles
whose annihilation in the halo is enhanced by the Sommerfeld effect. Since
we have at our disposal the full relative velocity distribution, we are
able to evaluate accurate fluxes for dark matter annihilations with an
arbitrary velocity dependence.

\section{Fluxes from dark matter annihilations}
\label{sec:flux}

With the results of the previous section we can evaluate the production
rate~\eq{avratevrel} for an arbitrary velocity dependent annihilation 
cross-section. In the following, we study gamma-ray fluxes from
the galactic center or a dSph satellite and synchrotron emission produced
by $e^\pm$ in the galactic magnetic field. We are motivated by the 
possibility that the recent observations of the rising positron fraction
at GeV energies could be due to DM annihilations. 
For this explanation to be viable, DM must dominantly
annihilate into leptons with a larger-than-thermal cross-section. If so,
there ought to be significant photon fluxes associated with these leptons: either
directly generated (via bremsstrahlung of charged particles or 
the decay of $\pi^0$) or the synchrotron radiation generated by the leptons
in the central galactic magnetic field. Such fluxes 
provide some of the most stringent constraints for these
scenarios~\cite{Bertone:2008xr}, since the large annihilation cross-section
is compounded with the increasing DM density in the center of the halo. 
In addition, the velocity dependence of the annihilation cross-section
enhances the emission from the center as pointed out in~\cite{Robertson:2009bh}
in the context of an approximate Jeans analysis. We here quantify this 
effect using the formalism developed above.

Let us start considering the photon fluxes. To simplify the discussion, 
suppose we are interested in photons of any energy and that the 
cross-section does not depend on energy. Then we have
\be
\frac{\diff \Gamma}{\diff V} = \frac{N_\gamma}{m_\chi^2} \langle \sigma v_\mathrm{rel} \rangle \rho_\chi^2,
\label{eq:lum_density}
\ee
where $N_\gamma$ is the number of photons that ultimately result from each 
annihilation, and we take $BR_i = 1$. 
%If we are concerned with photons seen by a telescope pointed in whatever direction, the differential volume containing relevant annihilations is
%\be
%\diff V = \diff A \,\diff l,
%\ee
%where $\diff A$ is the aperture {\bf (correct terminology??) } of the telescope and $\diff l$ is a line element inside the solid angle $\Delta \Omega$, centered normal to $\diff A$.
%Since nothing depends on the area element, 
The differential flux along a direction specified in general by two angles, 
$\psi$ and $\phi$, is
\be
\diff \Phi_\gamma = \diff l\,\frac{N_\gamma}{4 \pi m_\chi^2} \langle \sigma v_\mathrm{rel} \rangle \rho_\chi^2
\label{eq:lin_flux_density}
\ee
To find the total flux, we must integrate along the line-of-sight $l$ in all directions contained by the solid angle $\Delta \Omega$.
Any spatially-varying quantities in equation~\eq{lin_flux_density}, 
i.e. quantities that are functions of 
$\boldsymbol{r} = \boldsymbol{r}\left(l,\psi,\phi\right)$, must be taken 
into account in this integration.
The mass and photon yield per annihilation do not depend on location, 
and the density certainly does,
but what about the interaction rate $\langle \sigma v \rangle$\footnote{From
now on we suppress the subscript ``rel''}?
Integrating~\ref{eq:lin_flux_density}, the total flux from the galactic center is\cite{Bertone:2004pz}
\be
\label{eq:flux}
\Phi_\gamma(\Delta\Omega) = \frac{N_\gamma \langle \sigma v \rangle_\mathrm{MB}}{4 \pi m_\chi^2} \bar{J}\left(\Delta\Omega\right)\,\Delta\Omega,
\ee
where $\langle \sigma v \rangle_\mathrm{MB}$ is the interaction rate with the usual spatially-independent boost and the so-called ``$J$-factor'' is
\be
\label{J_factor}
J(\psi) = \int \mathrm{d}l\,\frac{\langle \sigma v \rangle(l)}{\langle \sigma v \rangle_\mathrm{MB}} \rho^2_\chi(l),
\ee
with the bar designating an average over the solid angle $\Delta\Omega$. We chose the $z$-axis
to point in the direction of the center of the halo, and we have assumed 
cylindrical symmetry. If we take the velocity distribution as Maxwell-Boltzmann
and spatially constant then $\langle \sigma v \rangle\left( l\right) = \langle \sigma v \rangle_\mathrm{MB}$ everywhere and
\be
J\left( \psi \right) \rightarrow J_\mathrm{MB}\left( \psi \right) = \int \mathrm{d}l\,\rho^2_\chi(l).
\ee
This is the usual calculation.

In general, though, $\langle \sigma v \rangle$ is location-dependent and must 
be kept inside the volume integral.
We work with the multiplicative change in the flux from using the Eddington equation or Jeans equation instead of Maxwell-Boltzmann:
\be
\label{eq:cal_F}
\mathcal{F} \equiv J/J_\mathrm{MB} = \Phi_\gamma / \Phi_{\gamma,\mathrm{MB}}.
\ee
DarkSUSY~\cite{darksusy} code rewritten in Python was used to calculate this 
quantity, given a profile model and Sommerfeld model.

We are also interested on the emission of synchrotron and inverse Compton (IC)
radiation in the galactic center. 
%If DM annihilations create electrons and positrons,
%the propogation of these particles will create radiation, especially near the galactic center where most of the annihilation occurs.
To find the synchrotron luminosity in the galaxy $L_\nu$, consider the 
energy distribution of electrons~\cite{Gondolo:2000pn,Bertone:2004pz}
\be
\label{eq:e_energy_dist}
\frac{\diff n_e}{\diff E} = \frac{\diff \Gamma}{\diff V}\frac{Y_e\left( > E \right)}{P_e\left( E \right)},
\ee
where $\diff \Gamma/\diff V$ is the local DM annihilation rate, $Y_e\left( > E \right)$ is the number of electrons and positrons created by each annihilation, and
\be
\label{eq:e_power_spectrum}
P_e\left( E \right) = \frac{2 e^4 B^2 E^2}{3 m_e^4 c^7}
\ee
is the electron power spectrum or energy loss rate.

The power radiated (as photons) at frequency~$\nu$ by an electron with energy~$E$ at galactocentric radius~$r$ is
\be
P_\gamma\left( \nu , E\right) = \frac{\sqrt{3} e^3}{m_e c^2} B\left( r\right) F\left(\frac{\nu}{\nu_c\left( E\right)}\right),
\ee
where $K_n$ is the modified Bessel function of order~$n$ and
\be
F\left(\frac{\nu}{\nu_c\left( E\right)}\right) = \frac{\nu}{\nu_c\left( E\right)} \int_0^\infty \diff y\, K_{5/3}\left( y\right).
\ee
Integrating over the electron energy $E$ and over the volume, we obtain the luminosity
\be
\label{eq:synch_lum_general}
L_\nu = \int \diff V\,\int_{m_e}^{\mchi} \diff E\,\frac{\diff n_e}{\diff E} P_\gamma\left( \nu ,E\right),
\ee
which becomes
\be
\label{eq:synch_lum_sub}
L_\nu = \frac{\sqrt{3} e^3}{m_e c^2} \int \diff V\,\frac{\diff \Gamma}{\diff V} B\left( r\right) \int_{m_e}^{\mchi} \diff E\,\frac{Y_e\left( > E\right)}{P_e\left( E\right)} F\left( \frac{\nu}{\nu_c \left( E\right)} \right) .
\ee
We focus on the spatially-dependent part of the integrand. The function $P\left( E \right)$ is proportional to $B^2$, so we have
\be
\label{eq:synch_lum_spatial}
L_\nu \sim \frac{4\pi}{\mchi^2}\int \diff r\, r^2\langle \sigma v \rangle \rho_\chi^2 B^3.
\ee
Most previous work takes the quantity $\langle \sigma v \rangle$ as spatially constant, removing it from the integrand.
To get an idea of the consequences of relaxing this approximation, we take a simple step-function as the model for the magnetic field.
The relevant quantity is then
\be
\label{eq:synch_lum_step}
L_\nu \sim \int_0^{r_B} \diff r\, r^2\langle \sigma v \rangle \rho_\chi^2,
\ee
where $r_B = 1\,\mathrm{kpc}$ is the radius at which we take the magnetic field as zero.

The multiplicative enhancement analogue to eq.~(\ref{J_factor}) is then:
%Similarly, the total annihilation rate within a spherical volume becomes
\be
\label{effective_Gamma}
\Gamma_\mathrm{ann} = 4 \pi \int_0^r \mathrm{d}r'\,r'^2\,\frac{\langle \sigma v \rangle (r')}{\langle \sigma v \rangle_0} \rho^2_\chi(r'),
\ee
where $r=r_B$ sets the size of the spherical volume where most of the emission
is generated. The same expression can be used to estimate the change
in the IC emission from the bulge, where the up-scattered
starlight and IR photons is most plentiful~\cite{Cirelli:2009vg}. Analogous to \eq{cal_F}, we focus on the quantity
\be
\label{eq:cal_G}
\mathcal{G} \equiv= \Gamma_\mathrm{ann}/\Gamma_{\mathrm{ann},\mathrm{MB}} = L_\nu/L_{\nu,\mathrm{MB}}.
\ee

Given a DM density profile and a particle physics model defining $\sigma v$,
we can calculate eqs.~(\ref{eq:cal_F})~and~(\ref{eq:cal_G}). We
choose the density profiles in Sec.~\ref{sec:dmprofile}, and consider
DM particles for which the cross-section times the flux is a decreasing
function of $v_\mathrm{rel}$.

\subsection{Sommerfeld enhancement}

The presence of a light force carrier mediating long range forces between
the dark matter particles causes a singular behaviour of the Feynman 
amplitude, which invalidates the partial-wave expansion $\sigma v_\mathrm{rel}
= a + b v_\mathrm{rel}^2 + \ldots$. This is the
basis of the non-perturbative \em{Sommerfeld} \em{enhancement}~\cite{Hisano:2004ds,Cirelli:2007xd,ArkaniHamed:2008qn,Pospelov:2008jd}. 
Although an accurate calculation of the enhancement requires the resummation
of processes with multiple exchanges of the mediator~\cite{iengo},
an analytic solution can be found by approximating the Yukawa potential
by the Hulth\'{e}n potential~\cite{Cassel:2009wt,Slatyer:2009vg,Feng:2010zp}.
The multiplicative change to the cross-section, called the Sommerfeld factor, is then
\be
S(v) = \frac{\pi \alpha_\chi}{v} \frac{\sinh\left(\frac{12\,v}{\pi \xi}\right)}{\cosh\left(\frac{12\,v}{\pi \xi}\right) - \cos\left(2 \pi \sqrt{\frac{6\,\alpha_\chi}{\pi^2 \xi} - \left(\frac{6\,v}{\pi^2 \xi}\right)^2}\right)}.
\ee
Here $\alpha_\chi$ is the coupling constant and $\xi$ 
is the ratio between the carrier mass and DM particle mass. 
This analytic expression is an excellent approximation to the numerical
calculation. It accurately reproduces the $\sigma v \sim 1/v$ behaviour 
that saturates to a constant $ \sim 1/v_{\mathrm{min}}$, with
$v_{\mathrm{min}} \approx \xi$, and also captures the resonant behaviour,
$\sigma v \sim 1/v^2$, for particular values of $\xi$~\cite{Feng:2010zp}.
The enhancement is then
\be
\frac{\langle \sigma\left(v\right) v\rangle\left(v\right)}{\langle \sigma v\rangle_0} = \int_0^\infty \mathrm{d}v\,S\left( v\right) P_\mathrm{rel}\left( v\right),
\label{eq:convolution}
\ee
where $\langle \sigma v\rangle_0$ is the annihilation rate before invoking the Sommerfeld enhancement.

To illustrate our calculations, we consider the 
Arkani-Hamed {\it et al.} model~\cite{ArkaniHamed:2008qn}, 
with $\alpha_\chi = 10^{-2}$ and $\xi$ between
$1.2\times 10^{-3}$ and $2\times 10^{-3}$. Within this range we find resonant
behaviour, as well as the standard $1/v$ regime that saturates 
at small velocities.

\begin{figure}[htb]
	\begin{center}
		\includegraphics[width=0.8\textwidth]{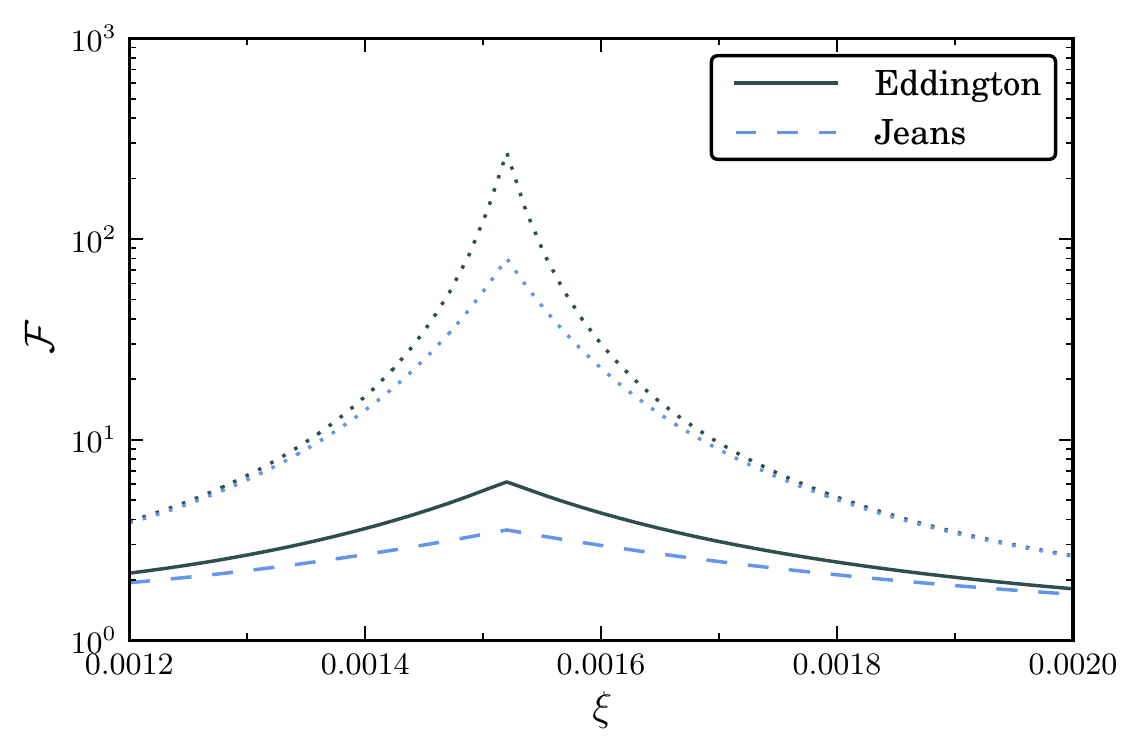}
	\end{center}
	\caption{$\mathcal{F}$ calculated for the galactic center using a NFW profile with (solid) and without (dotted) baryonic components.}
	\label{fig:gc_nfw_F}
\end{figure}

\begin{figure}[htb]
	\begin{center}
		\includegraphics[width=0.8\textwidth]{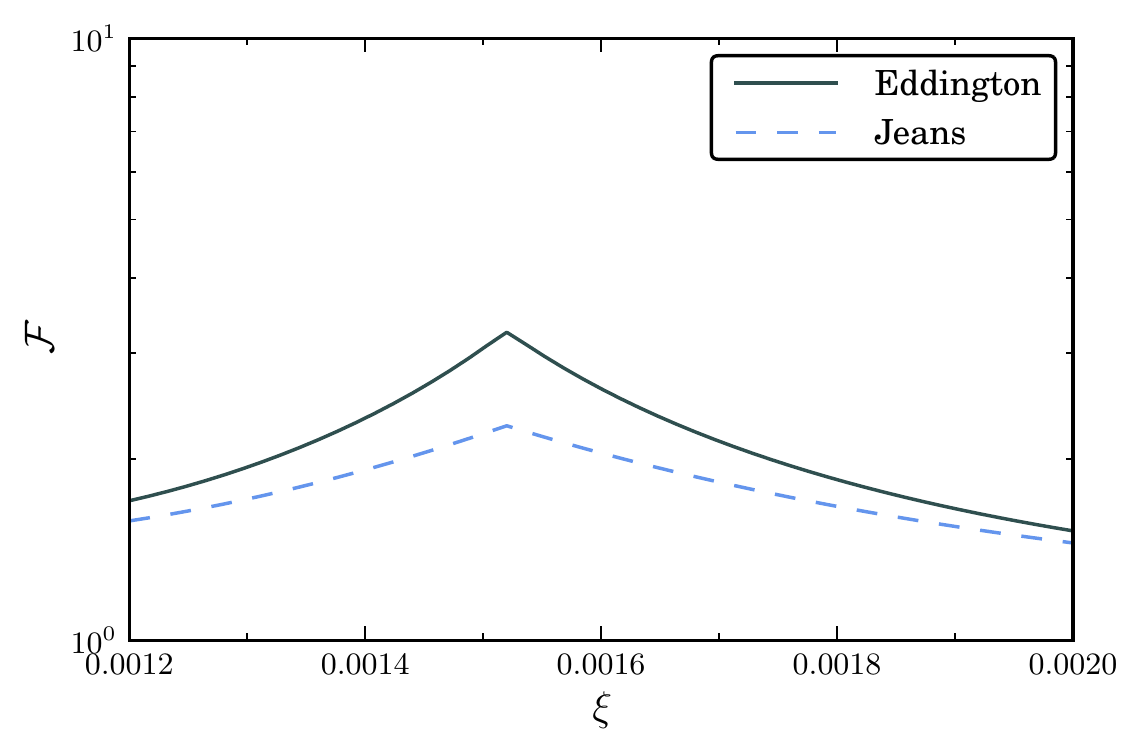}
	\end{center}
	\caption{$\mathcal{F}$ calculated for the galactic center using an Einasto profile with baryonic components.}
	\label{fig:gc_ein_F}
\end{figure}

\begin{figure}[htb]
	\begin{center}
		\includegraphics[width=0.8\textwidth]{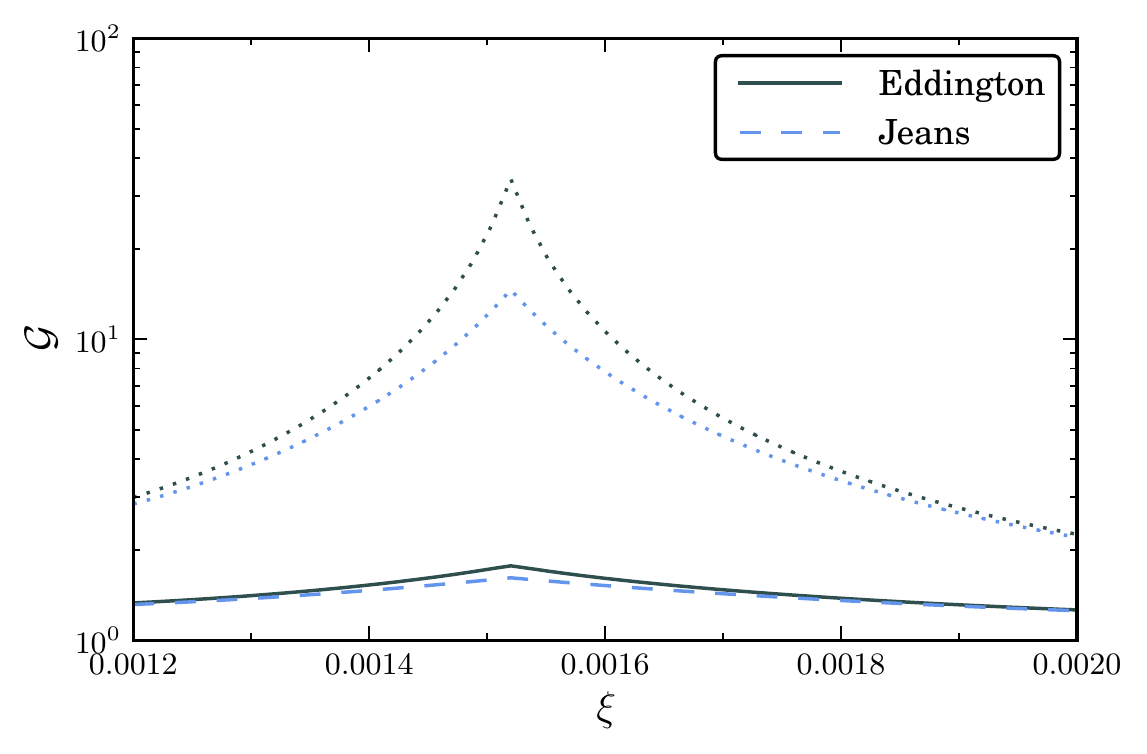}
	\end{center}
	\caption{$\mathcal{G}$ calculated using a NFW profile with (solid) and without (dotted) baryonic components.}
	\label{fig:gc_nfw_G}
\end{figure}

\begin{figure}[htb]
	\begin{center}
		\includegraphics[width=0.8\textwidth]{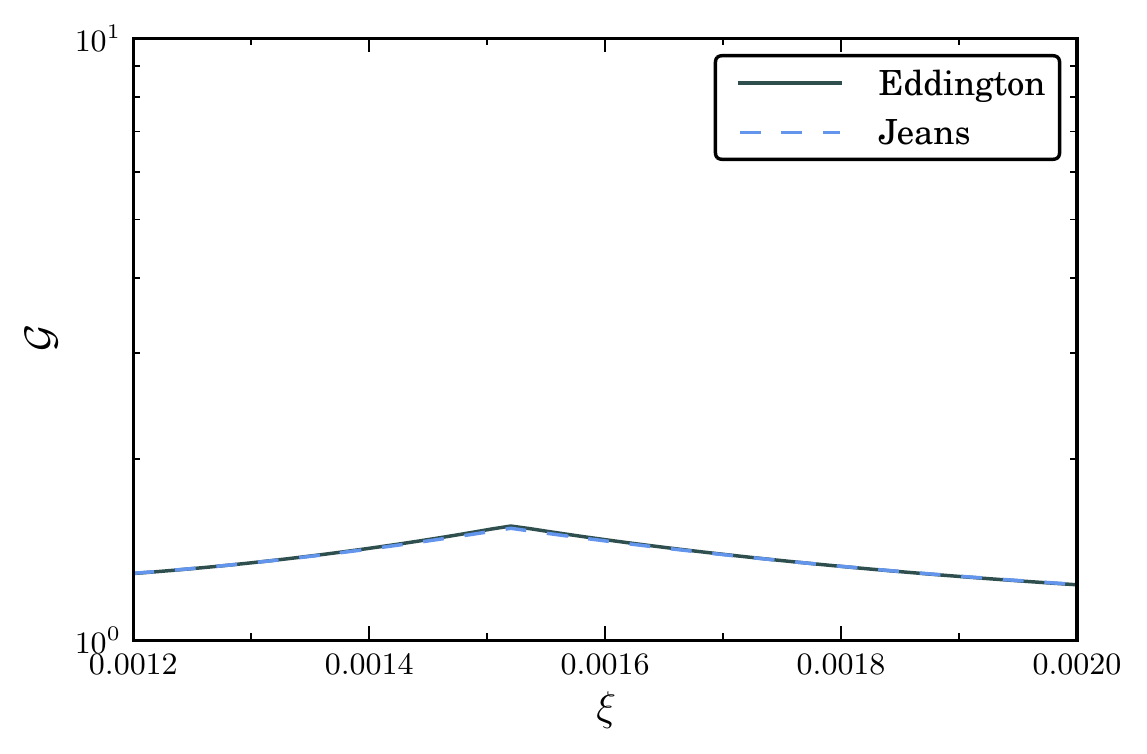}
	\end{center}
	\caption{$\mathcal{G}$ calculated using an Einasto profile with baryonic components.}
	\label{fig:gc_einasto_G}
\end{figure}

\begin{figure}[htb]
	\begin{center}
		\includegraphics[width=0.8\textwidth]{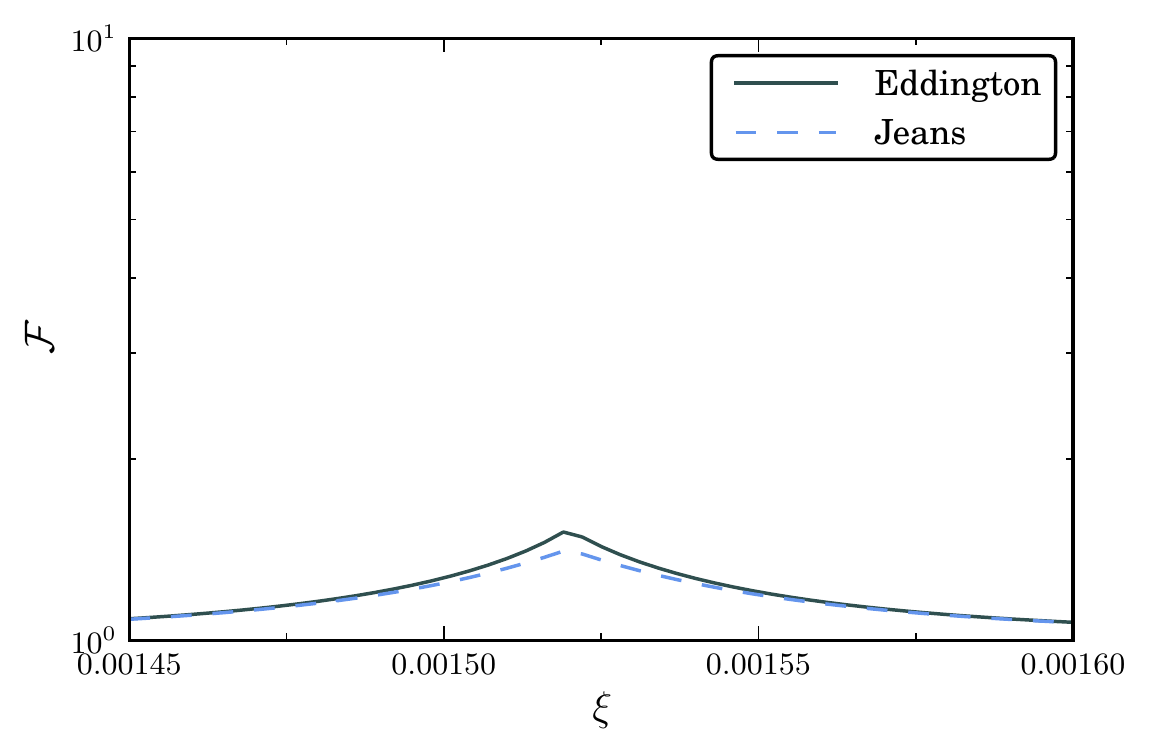}
	\end{center}
	\caption{$\mathcal{F}$ calculated for the Draco dwarf spheroidal using a Burkert profile.}
	\label{fig:draco_burkert_F}
\end{figure}

\begin{figure}[htb]
	\begin{center}
		\includegraphics[width=0.8\textwidth]{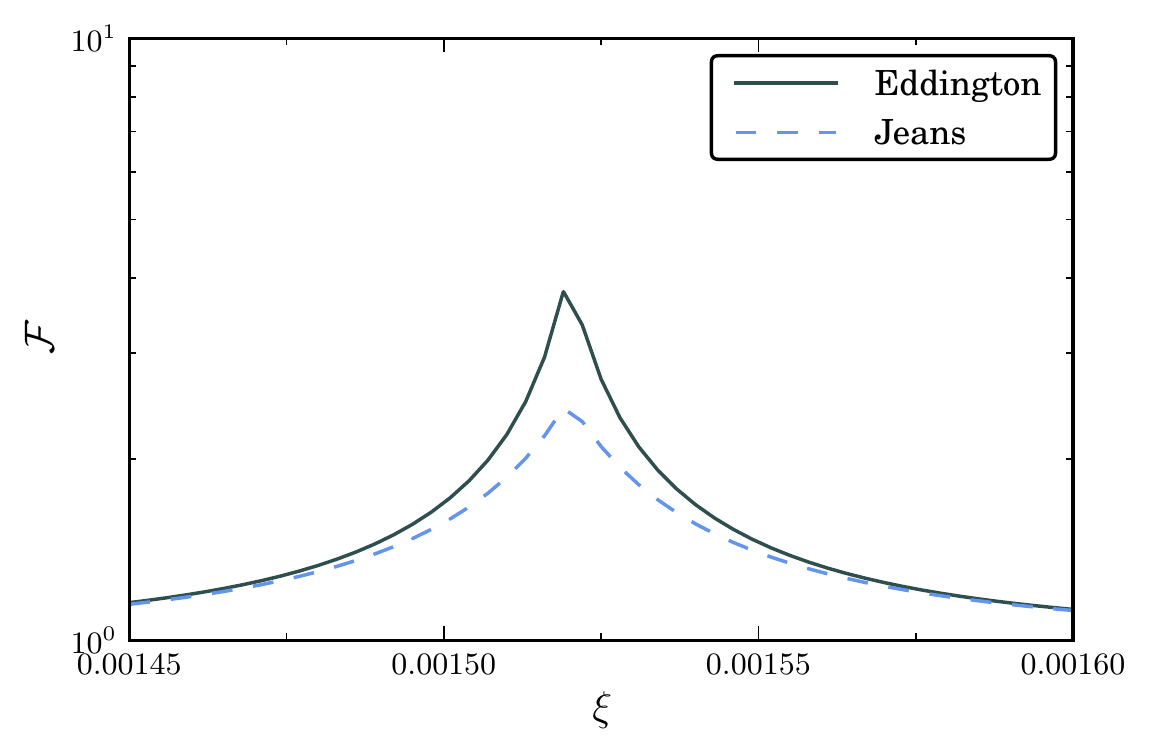}
	\end{center}
	\caption{$\mathcal{F}$ calculated for the Draco dwarf spheroidal using a NFW profile.}
	\label{fig:draco_nfw_F}
\end{figure}

\begin{table}
  \begin{center}
  \begin{tabular}{| c || c | c | c | c | c | c |}
    \hline
    \multicolumn{1}{|c||}{$\xi/10^{-3}$} & \multicolumn{3}{|c|}{$\mathcal{F}_\mathrm{DF} / \mathcal{F}_\mathrm{Jeans}$}  & \multicolumn{3}{|c|}{$\mathcal{F}_\mathrm{DF}$} \\
    \hline
    \multicolumn{1}{|c||}{ } & HO-NFW & NFW & Einasto & HO-NFW & NFW & Einasto \\
    \hline
                                  $1.2$  & 1.02 & 1.12 & 1.08 & 3.92 & 2.16 & 1.71 \\
                                  $1.3$  & 1.05 & 1.20 & 1.13 & 6.63 & 2.66 & 1.94 \\
                                  $1.45$ & 1.38 & 1.47 & 1.29 & 34.7 & 4.39 & 2.64 \\
                                  $1.51$ & 2.70 & 1.70 & 1.41 & 178. & 5.87 & 3.16 \\
                                  $1.6$  & 1.32 & 1.45 & 1.28 & 30.0 & 4.31 & 2.62 \\
                                  $1.8$  & 1.03 & 1.16 & 1.11 & 5.17 & 2.45 & 1.85 \\
                                  $2.0$  & 1.01 & 1.06 & 1.05 & 2.64 & 1.81 & 1.52 \\
    \hline
  \end{tabular}
  \caption{$J$-factor boosts for the galactic center. ``HO'' means halo-only; the other columns include baryonic components.}
  \end{center}
  \label{table:gc_J_enh}
\end{table}

\begin{table}
  \begin{center}
  \begin{tabular}{| c || c | c | c | c |}
    \hline
    \multicolumn{1}{|c||}{$\xi/10^{-3}$} & \multicolumn{2}{|c|}{$\mathcal{F}_\mathrm{DF} / \mathcal{F}_\mathrm{Jeans}$}  & \multicolumn{2}{|c|}{$\mathcal{F}_\mathrm{DF}$} \\
    \hline
    \multicolumn{1}{|c||}{ } & Burkert & NFW & Burkert & NFW \\
    \hline
                                  $1.48$ & 1.01 & 1.02 & 1.16 & 1.33 \\
                                  $1.5$  & 1.02 & 1.08 & 1.27 & 1.69 \\
                                  $1.52$ & 1.08 & 1.64 & 1.53 & 4.07 \\
                                  $1.54$ & 1.02 & 1.08 & 1.27 & 1.68 \\
                                  $1.56$ & 1.01 & 1.02 & 1.16 & 1.33 \\
    \hline
  \end{tabular}
  \caption{$J$-factor boosts for the Draco dwarf spheroidal.}
  \end{center}
  \label{table:draco_J_enh}
\end{table}

\section{Results and discussion}
\label{sec:discussion}
Calculating the relative velocity distribution,~\eq{fvrelp}, involves a
four-dimensional integration, which must be convolved according to
\eq{convolution} before performing the volume integration in eqs.~(\ref{eq:cal_F}) and~(\ref{eq:cal_G}). 
To avoid this numerically challenging route,
we generate a library of relative velocity distributions for each halo
profile. We pick $350$ ($250$) log-spaced locations between
$\rvir \leq r \leq 10^{-4} \,\rvir$ ($\rvir \leq r \leq 10^{-3}\,\rvir$)
in the Galaxy (dSph). At each location we sample $P_{\bm{r},\mathrm{rel}}$
for $10^3$ velocities between $0$ and the escape velocity at the distance
$r$ from the center\footnote{The collection of velocity distributions can
be downloaded from \url{http://www.physics.wustl.edu/ferrer/eddington/} .}.
We use these samples to evaluate \eq{convolution}.

Figures~\ref{fig:gc_nfw_F} and \ref{fig:gc_ein_F} show the change in flux $\mathcal{F}$, eq.~(\ref{eq:cal_F}), for the galactic center, modeled with 
a NFW halo and Einasto halo, respectively. A solid angle of $10^{-5}$sr is used, corresponding
to the typical angular resolution of the Large Area Telescope on board
of the Fermi satellite~\cite{fermi}.
Two different calculations of $\mathcal{F}$ are plotted: that of a Maxwell-Boltzmann distribution with variable dispersion 
found from the Jeans equation~\cite{Robertson:2009bh}, and 
our method based on Eddington's equation. 
Some observations are of note.
Using the Eddington equation generally grants predictions of stronger 
signals. This is because the distribution peak is at a lower 
velocity than in a Maxwell-Boltzmann distribution.
With any distribution, enhancements are generally lower when baryonic 
components are taken into account, which is obviously due to the extra 
mass that increases the velocity of the DM particles~\footnote{When the	cross-section increases with velocity, as in $p$-wave annihilation, baryons {\em increase} the flux.}
and it is also important to notice that the addition of these baryonic models lessen the difference between the Jeans analysis predictions and the 
Eddington predictions - significant deviation from a Maxwell-Boltzmann 
distribution occurs at a smaller radius (compare 
figures~\ref{fig:frel_nfwgal} and \ref{fig:frel_nfwgaldb}).
However, since much of the volume contributing to the line-of-sight 
observation is at the center, this deviation is still important. 

Table~\ref{table:gc_J_enh} shows some specific values of $\mathcal{F}$. 
Our Eddington based calculation shows that
when the variation in the velocity distribution of the DM particles is
taken into account, gamma-ray fluxes are a factor of 
$\sim 1 - 10$ {\em larger} than expected from the standard estimate. 
When this effect is estimated using the Jeans equation~\cite{Robertson:2009bh}
we typically recover fluxes within $\sim 20\%$ of the more accurate
Eddington prediction. In some particular cases, e.g. when the Sommerfeld
enhancement is resonant, the exact prediction can be more than $100$ times
larger than the usual rough estimate, and about three times larger than
what a Jeans analysis would bear.

Figures~\ref{fig:draco_burkert_F} and \ref{fig:draco_nfw_F} plot $\mathcal{F}$ for Draco, 
taken as an exemplary dwarf spheroidal, modeled using a Burkert and NFW profile, respectively. No baryonic components are included, since they
play a sub-dominant role in the dynamics of these galaxies.
Because of the distance, the solid angle consumes the entire object (it is nearly a point-source), so the entire volume contributes to the signal.
Since, as we have seen, a Maxwell-Boltzmann shaped distribution is accurate 
except at small radii, the flux enhancement is mostly due to the variation
of the dispersion and not the shape of the distribution, a deviation being 
significant only for a small range of Sommerfeld models. 
Table~\ref{table:draco_J_enh} gives some possible values of $\mathcal{F}$.

Figures~\ref{fig:gc_nfw_G} and \ref{fig:gc_einasto_G} plot $\mathcal{G}$ centered on the galaxy. The increase borne by our calculation 
is similar to that for dwarf spheroidals, $\sim 20-50\%$ since, again, the 
signal comes primarily from larger radii.

By this point it is prudent to stress that these calculations are 
dependent on the ``cut-off radius'', at and below which the DM density 
is taken to be constant.
DarkSUSY uses a default value of $10^{-5}\,\mathrm{kpc}$, which is much
smaller than the typical resolution of numerical N-body simulations. 
We take a much more conservative approach, keeping both density and
velocity distribution constant below 
$10^{-4}r_\mathrm{vir} = 2\times 10^{-2}\,\mathrm{kpc}$ 
($10^{-3}r_\mathrm{vir} = 3\times 10^{-3}\,\mathrm{kpc}$) 
for the Galaxy (Draco).
The velocity distribution, in any case, becomes more and more non-Maxwellian 
at smaller and smaller radii. In particular, if the same calculation took 
enhancements calculated from distributions down to 
$10^{-5}r_\mathrm{vir}$ or further, the impact of the spatial
dependence of the velocity distribution would be much more pronounced,
and our estimates are conservative in this respect.

Our calculation is self-consistent in the sense that our distribution function
correctly describes DM as a collisionless system confined by a known
gravitational potential. At the same time, our modelling of the galactic
is necessarily simplified and does not include all the components for a 
complete description. 

For instance, we have not allowed for any anisotropy
in the velocity distribution. The anisotropy is expected to be small close
to the center and increase in the outer region~\cite{Visbal:2012ta}. Since
most of the annihilations occur close to the center, isotropy seems a fair
assumption. However, the DF at a given position of the halo is sensitive
to all orbits with energy greater than the gravitational potential and
is affected, even in the central region, by the outer velocity 
anisotropy~\cite{mao}. Indeed, \cite{Mao:2012hf} finds the Eddington 
calculation to be an inferior fit to some galaxy-scale simulations.
These and other simulations (see, e.g. \cite{Baes:2002tw}) also suggest some 
anisotropy in the halo, which we plan to consider in future studies.

Also, we have focused on the smooth part of the halo, but small-scale
structure is seen in N-body simulations. Substructure was shown 
in~\cite{Slatyer:2011kg} to weaken somewhat the constraints from the
inner galaxy. This is so because cold subhalos are more likely to survive
in the outer galaxy and dominate the local signal, but they would have been 
tidally disrupted in the central regions. The presence of a dark disk
with low dispersion velocity, suggested by simulations that include baryons,
also seems to boost the local emission~\cite{Cholis:2010px}. However, the
dependence of the velocity distribution with the location in the halo was
not taken into account in these studies, and 
there are doubts about the presence of significant substructure in the 
local vicinity~\cite{Pieri:2009je}.

\paragraph{Non-spherical baryonic disk}
A limitation of our analysis is that it assumes that the system is spherical.
We have chosen a potential for the disk, following~\cite{Strigari:2009zb}, that mimics the gravitational pull of a
more realistic flattened system with an accuracy of $\sim 10\%$. Even though
Eddington's work has been extended to axisymmetric 
distributions~\cite{1962MNRAS.123..447L}, the formalism is far from trivial.
Nevertheless, we can estimate the effects of a flattened disk by means of
a simpler Jeans analysis, which does not need the full 
phase-space distribution function. The Jeans equation, \eq{jeans}, 
may also be expressed in cylindrical coordinates $(R,z)$:
\begin{align}
	{\bar{v}_R}^2(R,z) &= {\bar{v}_z}^2(R,z) = \frac{1}{\rho_\mathrm{DM}(R,z)} \int_z^\infty \mathrm{d}z\,\rho_\mathrm{DM}(R,z') \frac{\partial\Phi_\mathrm{tot}}{\partial z'}, \nn
	{\bar{v}_\phi}^2(R,z) &= {\bar{v}_R}^2 + \frac{R}{\rho_\mathrm{DM}} \frac{\partial \left( \rho_\mathrm{DM} {\bar{v}_R}^2(R,z) \right)}{\partial R} + R \frac{\partial \Phi_\mathrm{tot}}{\partial R}.
	\label{eq:jeanscyl}
\end{align}
Here, $\Phi_\mathrm{tot}$ is the total potential, with contributions from the
dark matter and the baryons.

To check the validity of the spherical disk model, 
we use~\eq{jeans} to calculate $\bar{v}^2$, 
and compare it to the analogous calculation, using~\eq{jeanscyl}, with the more 
realistic cylindrical disk model~\cite{binney}
\begin{equation}
\label{eq:sph_disk_rho}
\rho_\mathrm{d}(R,z) = \Sigma_\mathrm{d} e^{-R/R_\mathrm{d}} \left( \frac{\alpha_0}{2 z_0} e^{-|z|/z_0} + \frac{\alpha_1}{2 z_1} e^{-|z|/z_1} \right) ,
\end{equation}
taking $R_d = 4\,\mathrm{kpc}$, $z_0 = 0.3\,\mathrm{kpc}$, and 
$z_1 = 1\,\mathrm{kpc}$. 
The potential is
\begin{equation}
\Phi_\mathrm{d}(R,z) = -\frac{4 G \Sigma_\mathrm{d}}{R_\mathrm{d}} \int_{-\infty}^\infty \mathrm{d}z'\,\zeta (z') \int_0^\infty \mathrm{d}a\,\sin^{-1}\left(\frac{2 a}{\sqrt{+} + \sqrt{-}}\right) a K_0(a/R_\mathrm{d}),
\end{equation}
where $\zeta (z)$ is the expression inside the parentheses 
in~\eq{sph_disk_rho}, $\sqrt{\pm} \equiv \sqrt{(z-z')^2 + (a \pm R)^2}$, 
and $K_0$ is the modified Bessel function.

\begin{figure}[htb]
        \begin{center}
                \includegraphics[width=0.8\textwidth]{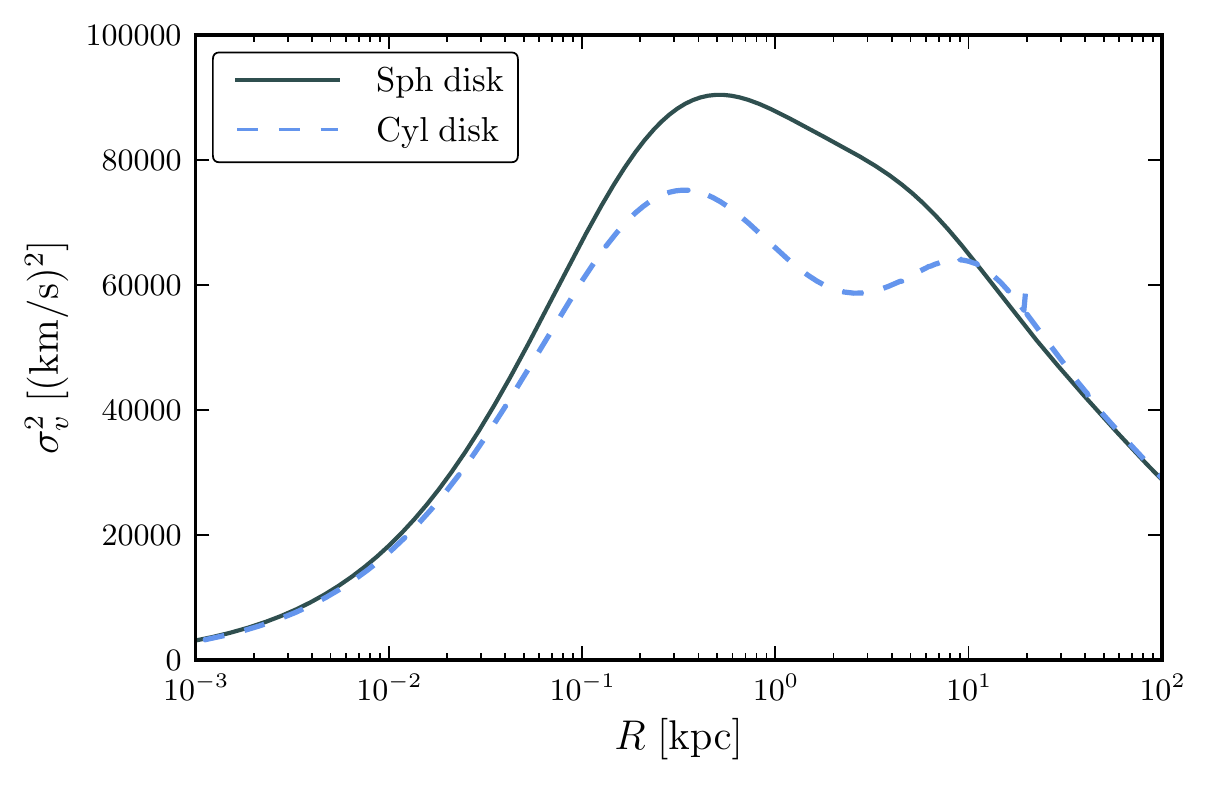}
        \end{center}
        \caption{Variance of the DM particle velocity as a function of
        the distance from the center on the galactic plane. A cylindrical disk model is
used in the first case; a spherical model is used in the latter. A NFW halo and a
baryonic bulge is used in both cases.}
        \label{fig:cyl_sph_plot}
\end{figure}

Figure~\ref{fig:cyl_sph_plot} shows the variance along the cylindrical radius for the two models.
Almost everywhere, the spherical disk model results in an overestimation of the variance by at most about 30\%
(so the dispersion is overestimated by at most about 15\% ). This means that predictions made of signals from Sommerfeld-enhanced annhilations will be conservative.

\section{Conclusion}

Using a self-consistent phase-space distribution for the galactic DM, we
have considered the annihilation of DM particles where $\sigma v$ is
an arbitrary function of the velocity.

We have found that, for models with Sommerfeld enhancement, annihilations in
the center of the halo are boosted, not only because of the larger density
predicted in N-body simulations, but also due to the smaller
average velocity of the DM particles. Including the latter effect raises
the annihilation fluxes from the galactic center, typically by a factor
of $\sim 1-10$, reaching $\gtrsim 100$ when resonances are present.

As a consequence, the stringent constraints from synchrotron and gamma-ray
emission on DM scenarios that attempt to explain the anomalies in the
positron fraction observed at GeV energies are further strengthened.

\acknowledgments
The authors are grateful to Jim Buckley, Wyn Evans and Michael Ogilvie 
for useful conversations. This
work was supported in part by the U.S. DOE under Contract
No. DE-FG02-91ER40628 and the NSF under Grant No. PHY-0855580. 
\appendix
\section{Relative velocity distribution for an ergodic system}
\label{sec:reldf}
Here, we write \eq{fvrelp} in terms of the individual phase-space 
distribution functions. Noting that \eq{fvrelp} is trivially zero
when $v_\mathrm{rel}=0$ or $v_\mathrm{rel} \geq \sqrt{8 \psit}$, 
and that $\tilde{f} \left(\epst\right)=0$
for $\epst \leq 0$, the integrand is non-zero when
\begin{align}
	0 \leq \tilde{v}_{\mathrm{cm}} \leq \sqrt{2 \psit}-\frac{\tilde{v}_\mathrm{rel}}{2} 
	\quad & \mathrm{and}
	\quad 0 \leq z \leq 1\nn
	&\mathrm{or} \nn
	\sqrt{2 \psit}-\frac{\tilde{v}_\mathrm{rel}}{2} \leq \tilde{v}_{\mathrm{cm}} \leq 
	\frac{\sqrt{8 \psit - \tilde{v}_\mathrm{rel}}}{2}
	\quad & \mathrm{and}
	\quad 0 \leq z \leq \frac{8 \psit - \tilde{v}_\mathrm{rel}^2 -
	4 \tilde{v}_\mathrm{cm}^2}{4 \tilde{v}_\mathrm{cm}\tilde{v}_\mathrm{rel}}.
\label{eq:limits} 
\end{align}
When $\tilde{v}_\mathrm{rel}=0$, \eq{fvrelp} is trivially zero. Otherwise,
\begin{align}
	P_{r, \mathrm{rel}}\left(v_\mathrm{rel}\right) =  
	\frac{2 \tilde{v}_\mathrm{rel}^2}{\pi^2 \rhot(r)^2}& 
	\left(\frac{\rvir}{G \mvir}
\right)^3\times   \nn 
&\left\{\int_0^{\sqrt{2 \psit} -\frac{\tilde{v}_{\mathrm{rel}}}{2}}{\diff 
	\tilde{v}_{\mathrm{cm}} 
\tilde{v}_{\mathrm{cm}}^2}  \int_{0}^1{\diff z } +
\int_{\frac{\sqrt{2 \psit} -\tilde{v}_{\mathrm{rel}}}{2}}^{\frac{\sqrt{8 \psit -
\tilde{v}_\mathrm{rel}^2}}{2}}
{\diff \tilde{v}_{\mathrm{cm}} 
\tilde{v}_{\mathrm{cm}}^2}  \int_{0}^{\frac{8 \psit - \tilde{v}_\mathrm{rel}^2 -
4 \tilde{v}_\mathrm{cm}^2}{4 \tilde{v}_\mathrm{cm}\tilde{v}_\mathrm{rel}}}{\diff z } \right\}
	\,\times \nn
	&	\tilde{f}\left( \psit(r)- \frac{\tilde{v}_{\mathrm{cm}}^2}{2} -
\frac{\tilde{v}_{\mathrm{rel}}^2}{8} - \frac{\tilde{v}_{\mathrm{cm}}\tilde{v}_{\mathrm{rel}} z}{2} \right)
\tilde{f} \left(\psit(r)- \frac{\tilde{v}_{\mathrm{cm}}^2}{2} -
\frac{\tilde{v}_{\mathrm{rel}}^2}{8} + \frac{\tilde{v}_{\mathrm{cm}}\tilde{v}_{\mathrm{rel}} z}{2}  \right). 
\label{eq:fvrel}
\end{align}

\bibliographystyle{JHEP}
\bibliography{vdm7}

\end{document}